\documentclass[a4paper,12pt]{article}

\usepackage{amsmath, amssymb, authblk, a4wide, bm, cancel, color, graphicx, subfig, enumerate, appendix, youngtab, cite, hyperref, bbold}

\newcommand{\Tr}[1]{\mathop{\rm Tr}\left[#1\right]}

\newcommand{\dimint}[2]{\int\mathrm{d}^{#1}#2\,}

\newcommand{\nbrack}[1]{\left(#1\right)}
\newcommand{\sbrack}[1]{\left[#1\right]}

\newcommand{\pfrac}[2]{\left(\frac{#1}{#2}\right)}
\newcommand{\sep}[1]{\quad\mbox{#1}\quad}
\newcommand{\brm}[1]{\bm{\mathrm{#1}}}
\def\fund{\tiny\Yvcentermath1\yng(1)}
\def\afund{\tiny\tilde{\Yvcentermath1\yng(1)}}
\def\sym{\tiny\Yvcentermath1\yng(2)}
\def\asym{\tiny\Yvcentermath1\yng(1,1)}
\def\widerow{\rule{0pt}{2.5ex}\rule[-1.5ex]{0pt}{0pt}}
\def\be{\begin{equation}}
\def\ee{\end{equation}}
\def\ba{\begin{eqnarray}}
\def\ea{\end{eqnarray}}
\def\cJ{{\cal J}}
\def\cL{{\cal L}}
\def\cN{{\cal N}}

\def\uno{\mbox{1 \kern-.59em {\rm l}}}

\numberwithin{equation}{section}
\numberwithin{figure}{section}
\numberwithin{table}{section}

\begin{document}

\title{{\normalsize IPPP/12/06, DCPT/12/12\hfill\mbox{}\hfill\mbox{}}\\
\vspace{2cm}
\Large{\textbf{Seiberg duality versus hidden local symmetry}}}
\author[1]{Steven Abel\thanks{\texttt{s.a.abel@durham.ac.uk}}}
\author[2]{James Barnard\thanks{\texttt{james.barnard@unimelb.edu.au}}}
\affil[1]{IPPP, Durham University, DH1 3LE, UK}
\affil[2]{School of Physics, University of Melbourne, Victoria 3010, Australia}
\date{}
\maketitle

\begin{abstract}
\baselineskip=15pt
\noindent It is widely believed that the emergent magnetic gauge symmetry of SQCD is analogous to a hidden local symmetry (HLS)\@.  We explore this idea in detail, deriving the entire (spontaneously broken) magnetic theory by applying the HLS formalism to spontaneously broken $SU(N)$ SQCD\@.  We deduce the K\"ahler potential in the HLS description, and show that gauge and flavour symmetry are smoothly restored along certain scaling directions in moduli space.  We propose that it is these symmetry restoring directions, associated with the $R$-symmetry of the theory, that allow full Seiberg duality.  Reconsidering the origin of the magnetic gauge bosons as the $\rho$-mesons of the electric theory, colour-flavour locking allows a simple determination of the parameter $a$.  Its value continuously interpolates between $a=2$ on the baryonic branch of moduli space -- corresponding to ``vector meson dominance" -- and $a=1$ on the mesonic branch.  Both limiting values are consistent with previous results in the literature.  The HLS formalism is further applied to $SO$ and $Sp$ groups, where the usual Seiberg duals are recovered, as well as adjoint SQCD.  Finally we discuss some possible future applications, including (naturally) the unitarisation of composite $W$ scattering, blended Higgs/technicolour models, real world QCD and non-supersymmetric dualities.
\end{abstract}

\newpage
\tableofcontents
\newpage 

\section{Introduction and conclusions}

Several ideas have been put forward for dealing with strongly coupled theories.  An early example, that works surprisingly well for QCD, is the notion of hidden local symmetry (HLS)\@.  Take a theory with flavour symmetry $G$ spontaneously broken to a subgroup $H$.  The strategy is to first use low energy theorems describing the associated Nambu-Goldstone bosons (NGBs) to construct an effective, nonlinear sigma model description on the manifold $G/H$ \cite{Coleman:1969sm,Callan:1969sn,Weinberg:1996kr}.  As it happens, this model is gauge equivalent to a linear model with flavour symmetry $G$ and a broken {\em gauge} symmetry $H$, thus providing an alternative effective description \cite{Bando:1984ej,Bando:1987br}.  The broken gauge symmetry is the HLS, in the sense that it was not present in the original theory but ``emerges'' in the low energy physics.

A somewhat more modern tool for tackling strongly coupled theories is Seiberg duality, applicable to certain $\cN=1$ supersymmetric gauge theories \cite{Seiberg:1994pq,Intriligator:1995au}.  In its default form, the duality links the low energy physics of $SU(N)$ SQCD with $N$ colours and $N+n$ flavours, with that of a second SQCD theory with $n$ colours, $N+n$ flavours of quark, some elementary singlets (identified as bound state mesons) and a Yukawa coupling between them all in the superpotential.  The duality also extends to $SO$ and $Sp$ gauge groups \cite{Intriligator:1995ne} as well as a veritable zoo of other, more intricate theories \cite{Kutasov:1995ve, Kutasov:1995np, Intriligator:1995ax, Pouliot:1995zc, Kutasov:1995ss, Pouliot:1995sk, Pouliot:1996zh, Brodie:1996vx, Brodie:1996xm, Abel:2009ty,Craig:2011tx,Craig:2011wj}.  The key feature of Seiberg duality is that it is a strong-weak duality which can, in certain circumstances, yield calculable results in strongly coupled theories.  In particular, choosing $N\ge2n$ renders the original, electric theory asymptotically free and the dual, magnetic theory IR free.

Despite the fact that Seiberg duality is well tested and well understood from a practical point of view, the question of {\em why} it exists has remained only partially answered.  Refs.~\cite{Harada:1999zj, Harada:2003jx} and, more recently, refs.~\cite{Komargodski:2010mc, Kitano:2011zk} reinterpreted the duality by proposing that the magnetic gauge group is in fact an HLS of the electric theory.  By analogy with QCD the magnetic gauge fields are then related to $\rho$-mesons of the electric theory.  Indeed, hints of this idea are already apparent in the flavour symmetry breaking pattern of SQCD: at a generic point in moduli space
\be
SU(N+n)_L\times SU(N+n)_R\longrightarrow SU(n)_L\times SU(n)_R\, ,
\ee
and the surviving non-Abelian factors clearly have the same form as the magnetic gauge group.

Previous investigations of this interpretation focused mainly on the phase structure of the theory, or on the Noether currents and comparison with QCD\@.  In this article we will instead return to the full, supersymmetric HLS formalism of refs.~\cite{Bando:1984cc,Bando:1987br}.  This approach yields many new results, all supporting the idea that Seiberg duality is just the way that HLS manifests itself in supersymmetry.

A difference between Seiberg duality and HLS that will be central to our discussion is that, whereas the former is a duality that exists between unbroken theories, the HLS is defined as a spontaneously broken symmetry.  The property of supersymmetric theories that allows us to reconcile this difference is that their potentials have enhanced complex flavour symmetries.  As a consequence their moduli spaces contain both NGBs {\emph {and}} quasi-NGBs, and there is a tendency for flavour symmetries to be broken by expectations of the latter.

One can therefore change the pattern of flavour symmetry breaking simply by moving around the moduli space.  It is then interesting to investigate what happens in the HLS description at points with enhanced flavour symmetry.  We will see that the previously broken HLS can be (partially) restored.  The newly massless gauge fields correspond to combinations of the NGBs that can no longer be considered NGBs at such points.  This is the general mechanism connecting HLS to Seiberg duality, and is the focus of section \ref{sec:HL}.

Applying the HLS formalism to SQCD (section \ref{sec:HQ}) provides an explicit realisation.  We initiate the procedure at a generic point in the electric theory's moduli space, where the flavour symmetry is maximally broken, and use the HLS formalism to recover the usual magnetic dual in a confined phase.  The magnetic superpotential is instrumental in this result.  It is required to avoid the double-counting of certain degrees of freedom, but it can also be considered as a UV completion that drives the breaking of the HLS (i.e.\ the magnetic gauge symmetry) via confinement.  We find it an appealing feature of SQCD that one can identify the dynamical mechanism that breaks the HLS\@. Equivalently, one can start from the magnetic theory, drive it into a confined phase via a meson expectation and recover the electric theory as the HLS description instead.

Regardless of which gauge group is taken to be the HLS, one can smoothly scale all symmetry breaking expectations to zero by moving along the quasi-NGB direction associated with spontaneously broken $R$-symmetry.  In this limit gauge and flavour symmetry are restored in both theories, thus recovering full Seiberg duality.

After re-establishing the HLS interpretation of Seiberg duality we derive several new results.  The mapping of electric $\rho$-mesons to magnetic gauge fields is immediate and explicit.  Comparing this result with a second derivation (using Noether currents and colour-flavour locking as in ref.~\cite{Komargodski:2010mc}) allows us to fix the value of the parameter $a$, analogous to that appearing in the chiral Lagrangian of real world QCD\@.  On a baryonic branch of SQCD we find $a=2$ (as in ref.~\cite{Komargodski:2010mc}) and on a mesonic branch $a=1$ (consistent with ref.~\cite{Kitano:2011zk}).

Electric quark mass terms are easily accommodated.  They reduce the size of the unbroken flavour symmetry leading to a higgsing of the magnetic gauge group.  We can also fix the duality scale, which would otherwise be a free parameter, at particular points in moduli space.  These results are presented in section \ref{sec:CH}.

In sections 5 and \ref{sec:VS} we extend the discussion in various novel directions; section 5 discusses what can be learnt by gauging $R$-symmetry, and in section \ref{sec:VS} we show that the HLS interpretation can be straightforwardly extended 
to the $SO$ and $Sp$ versions of Seiberg duality, as well as adjoint SQCD.

In addition to its theoretical interest, a better understanding of Seiberg duality opens up several areas of application.  Including operators charged under the magnetic gauge group in the duality's dictionary allows us to discuss dynamical processes.  This can lead to a semi-calculable description of the unitarisation of composite $W$ scattering.  By gauging the flavour symmetry one can also interpolate continuously between higgsing/technicolour descriptions of (supersymmetric) electroweak symmetry breaking.  Lessons learnt from applying the HLS formalism to SQCD may help understand the reason $a=2$ is selected in real world QCD\@.  Finally we (recklessly)  speculate that the whole procedure could be implemented as a systematic way of finding non-supersymmetric dualities.  These issues are discussed in section \ref{sec:AP}.

\section{Hidden local symmetry and SUSY\label{sec:HL}}

Consider a theory with a flavour symmetry $G$ broken to some subgroup $H$.  Low energy theorems tell us that the behaviour of the associated NGBs depends not on the specifics of the theory, but only on the symmetry breaking pattern $G\to H$.  Any effective Lagrangian that realises this provides a valid description of the underlying theory's IR physics \cite{Weinberg:1996kr}.

In non-supersymmetric theories a general approach is to realise the flavour symmetry nonlinearly, via a sigma model description on the manifold $G/H$ \cite{Coleman:1969sm,Callan:1969sn}.  It turns out that this description is gauge equivalent to one with a linearly realised flavour symmetry $G$ and a broken gauge symmetry $H$.  The gauge symmetry, which is not present originally, is said to emerge as a hidden local symmetry of the underlying theory.  Ref.~\cite{Bando:1987br} contains a comprehensive review of this idea and its application to supersymmetric theories.  We begin by briefly summarising this latter aspect.

SUSY tells us that each real NGB comes with two massless superpartners:  one real scalar and one Weyl fermion.  The extra light scalars can be considered a direct consequence of the holomorphy of the superpotential, which elevates real constants parameterising flavour transformations to complex ones.  At the superpotential level the original flavour symmetry is therefore enhanced to its complex extension $G^c$.  Typically this results in more symmetry generators being broken, hence more massless scalars.

While the genuine NGBs are coordinates for the real manifold $G/H$, the full set of massless scalars spans the larger, complex manifold $G^c/\hat{H}$.  Here $\hat{H}\supseteq H^c$ is the complex symmetry group preserved by the moduli space of the theory.  That it contains $H^c$ follows from the fact that generators of $H^c$ are constrained to be Hermitian, whereas those of $\hat{H}$ are not.  Thus a supersymmetric theory with flavour symmetry breaking $G\to H$ can be described by a sigma model on the manifold $G^c/\hat{H}$.

The NGBs are, of course, the usual massless scalars corresponding to the $G\to H$ part of the symmetry breaking.  Any other massless scalars are known as quasi-NGBs and are forbidden from getting mass terms only by SUSY\@.  The precise number of quasi-NGBs depends on how much bigger $G^c/\hat{H}$ is than $G/H$ and is given by
\be\label{eq:HLNM}
N_M=\dim{[G^c/\hat{H}]}-\dim{[G/H]}=\dim{[G]}+\dim{[H]}-\dim{[\hat{H}]}.
\ee
We count independent real dimensions such that $\dim{[G^c]}=2\dim{[G]}$ and so on.

Ref.~\cite{Bando:1987br} mainly studied the limiting case of $\hat{H}\simeq G\times H$, whereupon $N_M=0$ and all massless scalars are NGBs\@.  The other extreme is $\hat{H}=H^c$, whereupon $N_M=\dim{[G/H]}$ and there is a one to one correspondence between NGBs and superfields.  More generally, one can separate chiral superfields into P and M-types.  P-type (or pure) superfields have NGBs for both scalar components.  M-type (or mixed) superfields contain one NGB and one quasi-NGB\@.  Hence $N_M$ also gives the number of M-type superfields.

Whatever the value of $N_M$, one can define dimensionless superfields $\Pi^a$ to accommodate the normalised NGBs\@.  These are assembled into a chiral superfield matrix\footnote{Here, and henceforth, $x$ is used as shorthand for all superspace coordinates}
\be\label{eq:HLxiPi}
\xi(\Pi)=e^{\Pi(x)}\sep{where}\Pi(x)=\Pi^a(x)\hat{T}^a.
\ee
The $\hat{T}^a$ are the broken generators of $G^c$ and we have chosen a basis such that
\begin{align}
\Tr{\hat{T}^{a\dag}\hat{T}^b} & =\frac{1}{2}\delta^{ab} &
\Tr{\hat{S}^{\alpha\dag}\hat{S}^\beta} & =\frac{1}{2}\delta^{\alpha\beta} &
\Tr{\hat{T}^{a\dag}\hat{S}^\alpha} & =0
\end{align}
for generators $\hat{S}^\alpha$ of $\hat{H}$.  Note that these generators are not necessarily Hermitian for arbitrary complex groups.

The scalar components of $\xi$ are by definition elements of $G^c$, and provide standard representatives of each coset in the left coset space $G^c/\hat{H}$.  Acting on $\xi$ with a group element $g^\dag\in G$ does not usually preserve this parameterisation, but instead mixes in components involving the $\hat{S}^\alpha$ via
\be
g^\dag\xi(\Pi)=\xi(\Pi^\prime)\hat{h}^{-1}(\Pi,g)
\ee
for some element $\hat{h}^{-1}\in\hat{H}$ that depends on $\Pi$ and $g$ in some complicated way.  This means that $\xi$ transforms under flavour transformations according to
\be\label{eq:HLGNLS}
\xi(\Pi)\longrightarrow\xi(\Pi^\prime)=g\xi(\Pi)\hat{h}^{-1}(\Pi,g)
\ee
so the flavour symmetry is realised nonlinearly.  We could just as well have chosen $\xi$ to provide representatives of the {\em right} coset space $G^c/\hat{H}$ instead, whereupon $\hat{h}$ acts from the left and $g^\dag$ from the right in the above expression.

Before continuing, we briefly consider the expansion of the scalar component
\be\label{eq:HLxiPis}
\xi(\Pi)\supset e^{\kappa(x)}e^{i\pi(x)}
\ee
where $\kappa$ and $\pi$ are Hermitian scalar matrices.  Roughly speaking, the anti-Hermitian scalar part of $\Pi$ contains the NGBs and provides the phase factor $e^{i\pi}$.  This parameterises a nonlinear sigma model on $G/H$; which would have been constructed in a non-supersymmetric theory.  It satisfies the constraint $(e^{i\pi})^\dag(e^{i\pi})=\uno$ with the non-zero right hand side being a direct consequence of the symmetry breaking.  Meanwhile the Hermitian scalar parts of $\Pi$ provide $\kappa$, modifying this non-supersymmetric constraint to $\xi^\dag\xi=e^{2\kappa}$.  Therefore $\kappa$ parameterises fluctuations in the order parameters of the symmetry breaking.

That order parameters can appear as low energy degrees of freedom is a key feature of SUSY and, in principle, allows some of the symmetry breaking to be dialled down.  It occurs because said order parameters often arise from expectations of quasi-NGBs, leading to rich structure in the low energy theory.  This feature will be important when we come to discuss Seiberg duality as it enables the duality to be established for unbroken, not just broken, gauge symmetries.

Moving back to the task at hand we require any effective Lagrangian to be invariant under the nonlinear transformation \eqref{eq:HLGNLS}.  Building one is slightly trickier than in non-supersymmetric theories, mainly because $\hat{h}^\dag\neq\hat{h}^{-1}$ for an arbitrary complex group.  One proceeds by defining projection operators $\eta$ satisfying
\begin{align}\label{eq:HLeta}
\eta^\dag & =\eta & \eta^2 & =\eta & \hat{h}\eta & =\eta\hat{h}\eta
\end{align}
for every $\hat{h}\in\hat{H}$.  The first two expressions are generic features of such operators.  The third one states that the $\eta$ projected subspace (with $\eta$ acting from the right) is closed under $\hat{H}$, i.e.\
\be\label{eq:HLxiHeta}
\xi_\eta(\Pi^\prime)=\xi(\Pi^\prime)\eta=g\xi_\eta(\Pi)\hat{h}^{-1}_\eta(\Pi,g)\sep{where}\hat{h}^{-1}_\eta(\Pi,g)=\eta\hat{h}^{-1}(\Pi,g)\eta.
\ee
There is one projection operator for each $H$-irreducible block in $G$.

We are now able to write down a K\"ahler potential
\be\label{eq:HLK0}
K^S_\eta=v_\eta^2\ln{\det{\sbrack{\xi^\dag_\eta(\Pi^\dag)\xi_\eta(\Pi)}}}
\ee
for a real dimension 1 parameter $v_\eta$ associated with the scale of symmetry breaking in the underlying theory.  Because of the projection of $\xi_\eta$ and $\xi_\eta^\dagger$ it is not possible to split the holomorphic and anti-holomorphic factors in the determinant.\footnote{For this reason the identity is not included among the $\eta$'s as the resulting term in the K\"ahler potential does not contribute to the metric.}

This K\"ahler potential transforms according to
\begin{align}\label{eq:HLKSt}
K^{S\prime}_\eta & =v_\eta^2\ln{\det{\sbrack{\xi^\dag_\eta(\Pi^{\dag\prime})\xi_\eta(\Pi^\prime)}}} \nonumber\\
& =v_\eta^2\ln{\det{\sbrack{\hat{h}^{\dag-1}_\eta(\Pi^\dag,g^\dag)\xi^\dag_\eta(\Pi^\dag)g^\dag g\xi_\eta(\Pi)\hat{h}^{-1}_\eta(\Pi,g)}}} \nonumber\\
& =K^S_\eta+v_\eta^2\ln{\det{[\hat{h}^{\dag-1}_\eta(\Pi^\dag,g^\dag)]}}+v_\eta^2\ln{\det{[\hat{h}^{-1}_\eta(\Pi,g)]}}.
\end{align}
Since $\hat{h}^{-1}_\eta$ is a holomorphic function the last two terms have no D-term.  Therefore they do not contribute to the action which is consequently invariant.  Any linear combination of K\"ahler potentials of this form for different projection operators thus produces a suitable effective Lagrangian.  The resulting description is the expected nonlinear sigma model on the complex manifold $G^c/\hat{H}$.

We now turn to an HLS description, which is a linear description of the same system based on a theory with a flavour symmetry $G$ and a complex gauge symmetry $\hat{H}$.\footnote{Note that complex gauge symmetries are the norm in supersymmetric theories where gauge transformations are necessarily parameterised by chiral superfields. For example, in a theory with gauge group $H$ in the Wess-Zumino gauge, $H$ gauge transformations mix with SUSY transformations so that the full theory has a complexified gauge group $H^c$ in superspace.}  In the HLS theory, a chiral superfield $\xi$ is defined to live in a matrix representation of $G$ transforming as
\be
\xi(x)\longrightarrow g\xi(x)\hat{h}^{-1}(x).
\ee
This is essentially the same variable used above, although it is now considered to be an elementary chiral superfield rather than a function of $\Pi$.  The same projection operators \eqref{eq:HLeta} are introduced such that
\be
\xi_\eta(x)\longrightarrow g\xi_\eta(x)\hat{h}^{-1}_\eta(x)\sep{where}\hat{h}^{-1}_\eta(x)=\eta\hat{h}^{-1}(x)\eta
\ee
and we define an $\eta$ projected vector superfield $V_\eta=V(x)\eta(\hat{S}^\alpha+\hat{S}^{\alpha\dag})\eta$, transforming in the usual way
\be
e^{-V_\eta}\longrightarrow\hat{h}_\eta(x)e^{-V_\eta}\hat{h}^\dag_\eta(x)
\ee
under $\eta$ projected gauge transformations.

We then write down a K\"ahler potential
\be\label{eq:HLKVeta}
K^V_\eta=v_\eta^2\Tr{\pfrac{\xi_\eta^\dag(x)\xi_\eta(x)}{d_\eta^\dag(x)d_\eta(x)}e^{-V_\eta}+V_\eta}
\ee
i.e.\ a gauge theory with an FI-term $v_\eta^2V_\eta$ and $\hat{H}$ invariant auxiliary chiral superfield $d_\eta$.  One role of $d_\eta$ will be to ensure that any quasi-NGBs associated with the breaking of $U(1)$ symmetries remain unfixed by the vector superfield equations of motion.  A secondary role is clear from the fact that shifts in $d_\eta$ correspond to (complexified) gauge shifts in $V_\eta$, therefore restoring the ``radial" degrees of freedom that are absent in the non-linear sigma model.  We shall comment more on this later.

Currently there are no kinetic terms for the vector superfield.  It is an auxiliary degree of freedom and one can solve the equations of motion to find
\be\label{eq:HLVsolnS}
d_\eta^\dag(x)d_\eta(x)e^{V_\eta}=\xi_\eta^\dag(x)\xi_\eta(x).
\ee
(Note that the equations of motion still give $\xi_\eta^\dag\xi_\eta\propto\eta$ even if $\hat {H}$ does not contain a $U(1)$ subgroup such that the FI-term vanishes.)  Substituting \eqref{eq:HLVsolnS} back into eq.~\eqref{eq:HLKVeta}, ignoring constant terms
and a term proportional to $\ln{(d_\eta^\dag d_\eta)}$ (which vanish after the superspace integration) gives
\be
K^V_\eta=v_\eta^2\ln{\det{\sbrack{\xi_\eta^\dag(x)\xi_\eta(x)}}}=K^S_\eta.
\ee
Note that this expression is gauge invariant even though it has no $V_\eta$ because of the discussion around eq.~\eqref{eq:HLKSt}.  Hence any K\"ahler potential
\be\label{eq:HLKeta2}
K_\eta=(1-a_\eta)v_\eta^2\ln{\det{\sbrack{\xi_\eta^\dag(x)\xi_\eta(x)}}}+a_\eta v_\eta^2\Tr{\pfrac{\xi_\eta^\dag(x)\xi_\eta(x)}{d_\eta^\dag(x)d_\eta(x)}e^{-V_\eta}+V_\eta}
\ee
for an arbitrary, real constant $a_\eta$ reduces to $K^S_\eta$ after solving the vector superfield equations of motion.  This arbitrary constant relates the vector superfield mass to the gauge coupling $g_\eta$ via
$m_{V_\eta}^2=a_\eta g_\eta^2v_\eta^2$.  It is analogous to the $a$ parameter appearing in the chiral Lagrangian of regular QCD, where dynamics seems to fix its value to $2$.

Finally, observe that the transformation properties of $\xi$ mean that it can be written in the form
\be\label{eq:HLxisigma}
\xi(x)=e^{\Pi(x)}e^{\sigma(x)}\sep{where}\sigma(x)=\sigma^\alpha(x)\hat{S}^\alpha.
\ee
Fixing gauge such that $\sigma=0$ (the unitary gauge), $\xi$ becomes a function of $\Pi$ only and is identical to its counterpart in eq.~\eqref{eq:HLxiPi}.  However, flavour transformations do not respect this choice as
\be
\xi(\Pi(x))\longrightarrow g\xi(\Pi(x))=e^{\Pi^\prime(x)}e^{\sigma(x)}.
\ee
To maintain the gauge fixing they must be accompanied by gauge transformations
\be
\xi(\Pi(x))\longrightarrow g\xi(\Pi(x))\hat{h}^{-1}(\Pi(x),g)=\xi(\Pi^\prime(x))
\ee
exactly reproducing the nonlinear transformation properties of eq.~\eqref{eq:HLxiHeta}.

Despite the current lack of kinetic terms for the vector superfields many examples are known in which they develop dynamically, elevating the auxiliary vector superfields into true gauge fields.  Alternatively one might think of the HLS description without kinetic terms as being strongly coupled, with the kinetic terms being suppressed by the gauge coupling.  In SQCD the appropriate interpretation will depend on where one is on the moduli space.

That said, there is no symmetry stopping us adding gauge field kinetic terms to eq.~\eqref{eq:HLKeta2} by hand.  This is the approach usually taken for $\rho$-mesons in real world QCD, where it is anyway expected that kinetic terms are generated dynamically.

Note that, for the sake of pedagogy, we have been treating the whole of $\hat{H}$ democratically.  Often, different parts of $\hat{H}$ will have different properties.  For example, some of $\hat{H}$ may be anomalous, as will turn out to be the case in SQCD\@.  The HLS formalism remains applicable but the anomaly tells us that that part of the gauge symmetry cannot be restored.  In other cases kinetic terms may fail to arise dynamically for some components of $V_\eta$ which remain as auxiliary fields.  We shall continue to refer to the ``$\hat{H}$ gauge theory'', but bearing this in mind.

In summary, the two candidate low energy descriptions, a nonlinear sigma model on the complex manifold $G^c/\hat{H}$ and an $\hat{H}$ gauge theory with flavour symmetry $G$, are gauge equivalent to one another.  The $\hat{H}$ gauge symmetry arises as an HLS and there is one vector superfield for each projection operator satisfying eq.~\eqref{eq:HLeta} (or $H$-irreducible block of $G$).

\subsection{Restoring the gauge symmetry\label{sec:RGS}}

Many supersymmetric theories break their flavour symmetry via quasi-NGB expectations.  It is then possible to smoothly take a limit in which the symmetry is (partially) restored.  Here, we investigate in general terms how the HLS description behaves in this limit.  It will be an important component of our discussion of SQCD in the following section.

We begin by saying a little more about the auxiliary chiral superfield $d_\eta$.  In the Wess-Zumino gauge, the solution to the vector superfield equations of motion \eqref{eq:HLVsolnS} demands vacuum expectations
\be\label{eq:HLxiex}
\xi_\eta^\dag(x)\xi_\eta(x)=d_\eta^\dag(x)d_\eta(x)\eta
\ee
for the scalar components of $\xi$.  Gauge symmetry breaking is thus manifest for $d_\eta\neq0$: the scalar expectations provide a mass term for $V_\eta$.

Returning to eq.~\eqref{eq:HLxiPi}, suppose there is a broken generator $\hat{T}^1$ satisfying
\begin{align}\label{eq:HLT1def}
\hat{T}^1\eta & =\eta & [\hat{T}^1,\hat{T}^a]\eta & =0.
\end{align}
Equivalently, the restriction of $\hat{T}^1$ to the $\eta$ projected subspace generates a complexified $U(1)$ symmetry.  We can pull out the associated NGB superfield to write
\be\label{eq:HLPikappa}
\xi_\eta^\dag(\Pi)\xi_\eta(\Pi)=e^{2\bar{\kappa}_\eta(x)}\eta e^{\sum_{a\neq1}\Pi^{a\dag}(x)\hat{T}^{a\dag}}e^{\sum_{a\neq1}\Pi^a(x)\hat{T}^a}\eta
\ee
so the scale of $\xi_\eta^\dag\xi_\eta$ is able to fluctuate, as parameterised by the real scalar
\be\label{eq:HLkappadef}
\bar{\kappa}_\eta(x)=\frac{1}{2}\sbrack{\Pi^1(x)+\Pi^{1\dag}(x)}.
\ee
This is simply the superpartner of the NGB for the broken $U(1)$, and evidently a quasi-NGB of the theory.

As already stated, one job of $d_\eta$ is to ensure that the expectation of this quasi-NGB is not determined by the equations of motion.  To this end, eq.~\eqref{eq:HLxiex} fixes
$d_\eta^\dag d_\eta=e^{2\bar{\kappa}_\eta}$ in the Wess-Zumino gauge. 
The solution to the vector superfield equations of motion becomes
\be\label{eq:HLxiex2}
e^{2\bar{\kappa}_\eta(x)}e^{V_\eta}=\xi_\eta^\dag(x)\xi_\eta(x)
\ee
with the scalar components getting expectations $\xi_\eta^\dag\xi_\eta=e^{2\bar{\kappa}_\eta}\eta$ in the Wess-Zumino gauge as required.
Note that this issue is restricted to quasi-NGBs associated with complexified $U(1)$ symmetries in $\hat{H}$, since the FI-term in the K\"ahler potential is only relevant to them.  In the absence of $U(1)$ symmetries in $\hat{H}$ the vector superfield equations of motion give $\xi^\dag_\eta\xi_\eta\propto\eta$ but leave its overall scale unfixed.  

Our general strategy is now to define the HLS description at a point in moduli space where its gauge symmetry is broken, then use the quasi-NGB $\bar{\kappa}_\eta$ to move to where gauge symmetry is restored.  One cannot go directly to an HLS description with unbroken gauge symmetry as the sigma model description, from which the HLS description is derived, is not well defined at these points in moduli space.  The HLS description can, in fact, be considered to resolve the behaviour of the sigma model description as such points.

Eq.~\eqref{eq:HLxiex} implies that the expectation of $\xi_\eta$ vanishes and gauge symmetry is restored in the limit $d_\eta\to0$, i.e.\ $e^{\bar{\kappa}_\eta}\to0$.  However, the K\"ahler potential \eqref{eq:HLKeta2} blows up in this limit unless we simultaneously take $v_\eta\to0$.  Actually, this is precisely what we should do.  Recall from the discussion around eq.~\eqref{eq:HLxiPis} that $\bar{\kappa}_\eta$ parameterises fluctuations in the order parameter for the $\eta$ projected part of the symmetry breaking.  Since $v_\eta$ corresponds to the scale of this symmetry breaking, any scaling of $e^{\bar{\kappa}_\eta}$ is matched by a scaling of $v_\eta$.  Specifically, the combination $v_\eta e^{-\bar{\kappa}_\eta}$ remains fixed.

If we now define canonically normalised, dimensionful chiral superfields $q_\eta=\sqrt{a_\eta}v_\eta\xi_\eta/d_\eta$ the K\"ahler potential becomes
\be
K_\eta=(1-a_\eta)v_\eta^2\ln{\det{\sbrack{q_\eta^\dag(x)q_\eta(x)}}}+\Tr{q_\eta^\dag(x)q_\eta(x)e^{-V_\eta}+a_\eta v_\eta^2V_\eta}
\ee
up to terms that vanish after the superspace integral.  In the $e^{\bar{\kappa}_\eta}\to0$ limit with constant $v_\eta e^{-\bar{\kappa}_\eta}$ it dramatically simplifies to
\be
K_\eta=\Tr{q_\eta^\dag(x)q_\eta(x)e^{-V_\eta}}.
\ee
This K\"ahler potential is smooth everywhere and describes an unbroken gauge theory with massless matter, in which the $\eta$ projected part of $G\times\hat{H}$ is realised linearly.

The normalisation of $q$ is determined at the point in moduli space where the HLS description is first defined, then remains constant as the symmetry restoring limit is taken.  Hence many different normalisations lead to the same unbroken description.  We will see a manifestation of this effect in Seiberg duality: the duality scale $\mu$.  Also note that, as mentioned earlier, the massless degree of freedom that was absent from the sigma model, corresponding to the radial mode $v_\eta$,
can be associated with the $V_\eta$ field. Regardless of whether $V_\eta$ becomes a dynamical gauge field or remains as an auxiliary field, this degree of freedom passes to $q$ and $\tilde{q}$. (In the former case we can always go to the WZ gauge to make the bottom component of $V_\eta$ zero.)

Meanwhile, in the sigma model description, the K\"ahler metric based on eq.~\eqref{eq:HLK0} vanishes when $v_\eta\to0$.  Since part of the flavour symmetry is restored in this limit, it comes as no surprise to find that a description built around a broken flavour symmetry breaks down, and one has to revert to the underlying theory with restored flavour symmetry.

Indeed when $v_\eta\to0$ the $U(1)$ symmetry is restored so there is no longer a NGB, but the scalar partner of $\bar{\kappa}_\eta$ stays massless due to continuity of the moduli space.  The same thing happens to the NGBs of any other part of the flavour symmetry restored when $v_\eta\to0$.  Although the total number of massless degrees of freedom is unchanged the parameterisation used in eq.~\eqref{eq:HLxiPi} no longer makes sense.  Not all of the light states can be attached to a broken symmetry generator so the description is invalid.  

So we see that in both descriptions there are `new' massless degrees of freedom in the limit $v_\eta\to0$.  In the HLS description they are vector superfields, in the sigma model description they are former NGBs.  Comparing the two we therefore conjecture that HLS gauge fields emerge from NGBs that can no longer be considered such at points of enhanced symmetry in the underlying theory.

Using the equations of motion \eqref{eq:HLVsolnS} for the vector superfields we can actually be a little more specific.  They are easily rearranged to read
\be\label{eq:HLVsolnS2}
V_\eta=\ln{[e^{-2\bar{\kappa}_\eta(x)}\xi_\eta^\dag(x)\xi_\eta(x)]}=
\ln{[\eta e^{\sum_{a\neq1}\Pi^{a\dag}(x)\hat{T}^{a\dag}}e^{\sum_{a\neq1}\Pi^a(x)\hat{T}^a}\eta]}
\ee
upon substitution of eq.~\eqref{eq:HLPikappa}.  Expanding to leading order in the NGB superfields gives
\be
V_\eta\approx\ln{[\eta(\uno+\Pi^\dag)(\uno+\Pi)\eta]}\approx\eta(\Pi+\Pi^\dag)\eta+\eta\Pi^\dag\Pi\eta
\ee
where we use the fact that $\eta\uno\eta$ is simply the identity element of the $\eta$ projected subspace and the implied sum in $\Pi$ is understood.  Hence $V_\eta$ can indeed be related to the $\eta$ projection of the NGB superfields: precisely those whose scalar components are no longer NGBs when $v_\eta\to0$.

Finally, we can make a stronger claim for the vector superfields concerning the kinetic terms.  Returning to eq.~\eqref{eq:HLKeta2} we can rewrite the gauge coupling in the K\"ahler potential as
\be
K_\eta\supset a_\eta v_\eta^2\Tr{{\xi_\eta^\dag(x)\xi_\eta(x)e^{-(2\bar{\kappa}_\eta(x)+V_\eta)}}}
\ee
by factoring out the $\kappa_\eta$ part of $\xi_\eta$. As the dynamical degree of freedom $\kappa_\eta$ appears 
to mix with the trace of the vector superfield, it is plausible that the vector superfield becomes dynamical via a mixing with this quasi-NGB\@.  We will see more examples of the equivalence between vector superfields and quasi-NGBs later on. 

\section{Hidden local symmetry in SQCD\label{sec:HQ}}

Many of the abstract ideas of the previous section can be crystallised by considering the example of SQCD\@.  Indeed, we will provide further arguments that the magnetic dual can be interpreted as the HLS description as suggested in refs.~\cite{Komargodski:2010mc,Kitano:2011zk}.  To do so we will first derive the appropriate sigma model description, then go onto show that it is gauge equivalent to the usual magnetic theory.

We take an electric theory with $N$ colours and $N+n$ flavours.  The anomaly free flavour symmetry is
\be
G=SU(N+n)_L\times SU(N+n)_R\times U(1)_B\times U(1)_R
\ee
under which electric quarks $Q$ and $\tilde{Q}$ transform as per table \ref{tab:HQel}.  Gauge invariant meson and baryon operators
\begin{align}
M^i_j & =\tilde{Q}^i_\alpha Q^\alpha_j &
B_{j_1\ldots j_N} & =\epsilon_{\alpha_1\ldots\alpha_N}Q^{\alpha_1}_{j_1}\ldots Q^{\alpha_N}_{j_N} & \tilde{B}^{i_1\ldots i_N} & =\epsilon^{\alpha_1\ldots\alpha_N}\tilde{Q}_{\alpha_1}^{i_1}\ldots\tilde{Q}_{\alpha_N}^{i_N}
\end{align}
parameterise the theory's moduli space.  At a generic point they pick up expectations
\begin{align}\label{eq:HQelvevs}
M & ={\rm diag}\nbrack{\tilde{v}_1v_1,\ldots,\tilde{v}_Nv_N,0,\ldots,0} &
B_{1\ldots N} & =v_1\ldots v_N &
\tilde{B}^{1\ldots N} & =\tilde{v}_1\ldots\tilde{v}_N
\end{align}
up to symmetry transformations.  D-flatness demands that the difference $|v_i|^2-|\tilde{v}_i|^2$ is a constant but the parameters are otherwise free.

For non-zero $v$'s and $\tilde{v}$'s the gauge symmetry is completely broken.  Since the electric quark matrices are rank $N$, the flavour symmetry breaking is limited to
\be\label{eq:HQsympatt}
H=SU(n)_L\times SU(n)_R\times U(1)_{B^\prime}\times U(1)_{R^\prime}
\ee
where the unbroken $U(1)$ symmetries are a mixture of the original ones with $SU(N+n)_L\times SU(N+n)_R$ transformations.  The order parameters are conveniently organised by defining
\begin{align}
v & =|B_{1\ldots N}|^{1/N} &
\tilde{v} & =|\tilde{B}^{1\ldots N}|^{1/N}
\end{align}
and there is a constraint on the moduli space
\be
B_{1\ldots N}\tilde{B}^{1\ldots N}-{\rm det}_N(M)=0.
\ee

\begin{table}[!tb]
\be
\begin{array}{|c|c|cccc|}\hline
\widerow & SU(N) & SU(N+n)_L & SU(N+n)_R & U(1)_B & U(1)_R \\\hline
\widerow Q & \fund & \afund & {\bm1} & \phantom{-}1/N\phantom{-} & n/(N+n) \\
\widerow \tilde{Q} & \afund & {\bm1} & \fund & -1/N\phantom{-} & n/(N+n) \\\hline
\end{array} \nonumber
\ee
\caption{The matter content of the electric theory.  The first $SU(N)$ is the gauge group.\label{tab:HQel}}
\end{table}

Let us now look at one sector of the symmetry breaking in detail, e.g.\ $SU(N+n)_L\to SU(n)_L$.  Without loss of generality we consider this to be triggered by an expectation of the $N\times(N+n)$ quark matrix
\be\label{eq:HQQvev}
Q=\begin{pmatrix} {\bm v} & 0 \end{pmatrix}\sep{where} {\bm v}={\rm diag}(v_1,\ldots,v_N).
\ee
The broken and unbroken generators acting (somewhat confusingly) on the right are complex matrices with the forms
\begin{align}\label{eq:HQgens}
\hat{T}_L & =\bordermatrix{& N & n \cr N & \hat{T}_{L,N}+n\uno & \hat{T}_u \cr n & 0 & -N\uno} & 
\hat{S}_L & =\bordermatrix{& N & n \cr N & 0 & 0 \cr n & \hat{S}_l & \hat{S}_{L,n}}
\end{align}
up to unimportant normalisation factors.  Both $\hat{T}_{L,N}$ and $\hat{S}_{L,n}$ are traceless.

An $SU(n)^c$ subgroup is evidently generated by $\hat{S}_{L,n}$ but there remain $2Nn$ additional unbroken generators.  Therefore
\be
\dim{[\hat{H}_L]}=2(n^2-1)+2Nn
\ee
and eq.~\eqref{eq:HLNM} tells us that there are $N^2$ quasi-NGBs\@.  In other words there are $N^2$ M-type and $Nn$ P-type superfields associated with the $SU(N+n)_L$ factor, saturating the degrees of freedom available in $Q$.  Similar reasoning applies for the $SU(N+n)_R$ factor where
\begin{align}\label{eq:HQTRdef}
\hat{T}_R & =\begin{pmatrix} \hat{T}_{R,N}-n\uno & 0 \\ \hat{T}_l & N\uno \end{pmatrix} &
\hat{S}_R & =\begin{pmatrix} 0 & \hat{S}_u \\ 0 & \hat{S}_{R,n} \end{pmatrix}
\end{align}
for traceless matrices $\hat{T}_{R,N}$ and $\hat{S}_{R,n}$ acting on the left (for consistent confusion).

In addition to these generators there are identity matrices from each of the two $U(1)$ factors.  Only a linear combination of the original baryon number generator with $\hat{T}_L$ and $\hat{T}_R$ is broken.  It can be absorbed into the existing generators by redefining
\begin{align}\label{eq:HQTSL}
\hat{T}_L & =\begin{pmatrix} \hat{T}_{L,N}+\uno & \hat{T}_u \\ 0 & 0 \end{pmatrix} &
\hat{T}_R & =\begin{pmatrix} \hat{T}_{R,N}-\uno & 0 \\ \hat{T}_l & 0 \end{pmatrix}
\end{align}
and
\begin{align}\label{eq:HQTSR}
\hat{S}_L & =\begin{pmatrix} 0 & 0 \\ \hat{S}_l & \hat{S}_{L,n}+\uno \end{pmatrix} &
\hat{S}_R & =\begin{pmatrix} 0 & \hat{S}_u \\ 0 & \hat{S}_{R,n}-\uno \end{pmatrix} &
\hat{S}_{B^\prime} & =\begin{pmatrix} 0 & 0 \\ 0 & \uno \end{pmatrix}
\end{align}
with the difference in sign
for the $\uno$ component arising from $Q$ and $\tilde{Q}$ having equal and opposite baryon number.  On the other hand, we include the broken $R$-symmetry generator in its entirety.  Unbroken symmetry transformations therefore have the forms
\begin{align}\label{eq:HQhLRdef}
\hat{h}_L & =\begin{pmatrix} \uno & 0 \\ \hat{h}_{L,l} & \hat{h}_{L,n} \end{pmatrix} &
\hat{h}_R & =\begin{pmatrix} \uno & \hat{h}_{R,u} \\ 0 & \hat{h}_{R,n} \end{pmatrix} &
\end{align}
where $\det{(\hat{h}_{L,n})}=\det{(\hat{h}_{R,n})}=1$, in addition to a separate $U(1)_{B^\prime}$ transformation.
 
Including the full $R$-symmetry generator in the $\hat{T}$'s enables us to avoid complications inherent to gauged $R$-symmetries (discussed in section \ref{sec:GR}).  However, it also leads to different representatives for the $G^c/\hat{H}$ coset space relative to eq.~\eqref{eq:HLxiPi}, as we do not remove all contributions from the $\hat{H}$ generators.  One can think of this as leaving more gauge redundancy in the $\xi$'s than is usual in the HLS formalism.  Even though it isn't explicitly part of the HLS, we will still see gauge-like properties in the $R$-symmetry sector of the HLS description.

\subsubsection*{Standard coset description}

\noindent At this point we have to decide whether we want the sigma model description's variables \eqref{eq:HLxiPi} to live in left or right cosets of $G^c/\hat{H}$.  The obvious choice is for $\hat{h}_L$ to act on the right and $\hat{h}_R$ on the left (and vice-versa for $g_L$ and $g_R$), mirroring the original flavour transformations of the quarks.  It is then possible to find a unique projection operator satisfying eq.~\eqref{eq:HLeta} for $\hat{h}_L$, and $\eta\hat{h}_R=\eta\hat{h}_R\eta$ for $\hat{h}_R$:
\be\label{eq:HQetadef}
\eta=\begin{pmatrix} 0 & 0 \\ 0 & \uno \end{pmatrix}.
\ee
The action of $\hat{h}_L$ and $\hat{h}_R$ on the $\eta$ projected subspace
\begin{align}
\eta\hat{h}_L\eta & =\hat{h}_{L,n} & \eta\hat{h}_R\eta & =\hat{h}_{R,n}
\end{align}
is simply an $SU(n)^c_L\times SU(n)^c_R\times U(1)_{B^\prime}^c$ transformation.

Using eqs.~\eqref{eq:HLxiPi}, \eqref{eq:HQTSL} and \eqref{eq:HQTSR} we can then define chiral superfields
\begin{align}\label{eq:HQxidef}
\xi & =e^{\kappa_R}\begin{pmatrix} \xi_N & \xi_u \\ 0 & \uno \end{pmatrix} &
\tilde{\xi} & =e^{\kappa_R}\begin{pmatrix} \tilde{\xi}_N & 0 \\ \tilde{\xi}_l & \uno \end{pmatrix}
\end{align}
where $\det{(\tilde{\xi}_N\xi_N)}=1$, and the independent superfield $\kappa_R$ comes from the broken $U(1)_R$ generator.  Applying the projection operator we find a low energy sigma model description in terms of chiral superfields $\xi_\eta=\xi\eta$ and $\tilde{\xi}_\eta=\eta\tilde{\xi}$ transforming as
\begin{align}\label{HQqxi}
\xi_\eta=e^{\kappa_R}\begin{pmatrix} \xi_u \\ \uno \end{pmatrix} & \longrightarrow g_L\xi_\eta\hat{h}_{L,n}^{-1} &
\tilde{\xi}_\eta=e^{\kappa_R}\begin{pmatrix} \tilde{\xi}_l & \uno \end{pmatrix} & \longrightarrow\hat{h}_{R,n}\tilde{\xi}_\eta g_R^\dag
\end{align}
and with equal and opposite charge under $U(1)_{B^\prime}$.  The nonlinear dependence of $\hat{h}_{L,n}^{-1}$ on $\Pi$ and $g_L$ is understood.

Both $\xi_u$ and $\tilde{\xi}_l$ contain $Nn$ chiral superfield degrees of freedom generated by off diagonal components of the broken generators.  These are the P-type superfields of the flavour symmetry breaking.  Meanwhile $\kappa_R$ is an M-type superfield.  The K\"ahler potential for this description follows straight from eq.~\eqref{eq:HLK0} and is
\be\label{eq:HQKsigma}
K^S=\Tr{v^2\ln{(\xi_\eta^\dag\xi_\eta)}+\tilde{v}^2\ln{(\tilde{\xi}_\eta\tilde{\xi}_\eta^\dag)}}.
\ee

\subsubsection*{Flipped coset description}

\noindent Alternatively we can choose $\hat{h}_L$ to act on the left and $\hat{h}_R$ on the right (and versa-vice for $g_L$ and $g_R$).  Eq.~\eqref{eq:HQxidef} is mostly unchanged but the projection operator should now satisfy $\eta^\prime\hat{h}_L=\eta^\prime\hat{h}_L\eta^\prime$ and $\hat{h}_R\eta^\prime=\eta^\prime\hat{h}_R\eta^\prime$; the unique solution being
\be
\eta^\prime=\begin{pmatrix} \uno & 0 \\ 0 & 0 \end{pmatrix}.
\ee
This is actually a special choice because it satisfies $\eta^\prime\hat{h}_L=\eta^\prime\hat{h}_R=\eta^\prime$.  Hence the $\eta^\prime$ projected subspace is invariant under $\hat{H}$, not just closed.
The unbroken baryon number symmetry is also projected out.  This already suggests that flipping the cosets should result in a description of quasi-NGBs acting as order parameters, which are invariant under $\hat{H}$ by definition.

We are thus able to define a second set of chiral superfields $\xi_\eta^\prime=\eta^\prime\xi$ and $\tilde{\xi}_\eta^\prime=\tilde{\xi}\eta^\prime$ with transformation properties
\begin{align}
\xi_\eta^\prime=e^{\kappa_R}\begin{pmatrix} \xi_N & \xi_u \end{pmatrix} 
& \longrightarrow\xi_\eta^\prime g_L^\dag &
\tilde{\xi}_\eta^\prime = e^{\kappa_R}
\begin{pmatrix} \tilde{\xi}_N \\ \tilde{\xi}_l \end{pmatrix}  & \longrightarrow g_R\tilde{\xi}_\eta^\prime
\end{align}
which are conveniently assembled into a single variable
\be\label{eq:HQMdef}
\tilde{\xi}_\eta^\prime\xi_\eta^\prime=e^{2\kappa_R}\begin{pmatrix} \tilde{\xi}_N\xi_N & \tilde{\xi}_N\xi_u \\ \tilde{\xi}_l\xi_N & \tilde{\xi}_l\xi_u \end{pmatrix}\longrightarrow g_R[\tilde{\xi}_\eta^\prime\xi_\eta^\prime]g_L^\dag.
\ee
Due to its invariance under $\hat{H}$, the K\"ahler potential for $\tilde{\xi}_\eta^\prime\xi_\eta^\prime$ is constrained only by the flavour symmetry.  Any real function invariant under $SU(N+n)_L\times SU(N+n)_R\times U(1)_B\times U(1)_R$ will do.

Note that this superfield contains the remaining $N^2$ M-type superfields, thus encapsulating the other low energy quasi-NGB degrees of freedom (although there are $N^2$ from each half of the flavour symmetry breaking, half of them are eaten by the electric gauge fields and one accounted for by $\kappa_R$).  It also contains a copy of the superfield degrees of freedom that are already present in $\xi_\eta$ and $\tilde{\xi}_\eta$, although now of course transforming in a different way.  Thus we must ensure that these degrees of freedom are included only once if we wish to use both standard and flipped coset descriptions at the same time.

\subsection{Rediscovering the magnetic theory\label{sec:RM}}

We now seek an HLS description.  Following section \ref{sec:HL} we consider a theory based on an $SU(n)_L\times SU(n)_R\times U(1)_{B^\prime}$ gauge group.  It is immediately apparent that (individually at least) both $SU(n)$ factors have cubic anomalies for the anticipated matter content.  In practise there is nothing stopping us implementing an anomalous gauge symmetry in an effective theory, as long as its gauge fields are massive (which of course they are by construction in the HLS description).  Clearly a limit in which {\em all} gauge fields become massless cannot exist.  Nonetheless, the approach still gives a valid description of the low energy physics and, moreover, there is nothing forbidding limits in which the gauge fields of an anomaly free subgroup become massless.

\begin{table}[!tb]
\be
\begin{array}{|c|cc|cc|c|cc|}\hline
\widerow & SU(n)_L & SU(n)_R & SU(n) & SU(n)^\prime & U(1)_{B^\prime} & SU(N+n)_L & SU(N+n)_R \\\hline
\widerow \xi_\eta & \afund & {\bm1} & \afund & \afund & +1 & \fund & {\bm1} \\
\widerow \tilde{\xi}_\eta & {\bm1} & \fund & \fund & \afund & -1 & {\bm1} & \afund \\\hline
\end{array} \nonumber
\ee
\caption{The matter content of the standard coset HLS description.  The first two $SU(n)$ factors give one description of the gauge group, the second two define an alternative linear combination.  The final two factors are flavour symmetries\label{tab:HQxi}}
\end{table}

Guided by eq.~\eqref{eq:HLKeta2} we therefore write down a K\"ahler potential
\begin{align}
K= & \Tr{(1-a)v^2\ln{(\xi_\eta^\dag\xi_\eta)}+(1-\tilde{a})\tilde{v}^2\ln{(\tilde{\xi}_\eta\tilde{\xi}_\eta^\dag})}+ \nonumber\\
& \Tr{av^2\nbrack{\pfrac{\xi_\eta^\dag\xi_\eta}{d_\eta^\dag d_\eta}e^{V_{B^\prime}-V_L}-V_{B^\prime}}+\tilde{a}\tilde{v}^2\nbrack{\pfrac{\tilde{\xi}_\eta\tilde{\xi}_\eta^\dag}{d_\eta^\dag d_\eta}e^{V_R-V_{B^\prime}}+V_{B^\prime}}}
\end{align}
for the standard coset description.  All superfields are functions of superspace coordinates and transform as in table \ref{tab:HQxi}.  The vector superfields $V_L$ and $V_R$ are constructed from the $\eta$ projected generators $\eta(\hat{S}_L+\hat{S}_L^\dag)\eta$ and $\eta(\hat{S}_R+\hat{S}_R^\dag)\eta$ respectively.
The FI-terms pick out $V_{B^\prime}$, the $U(1)_{B^\prime}$ vector superfield being the only one with non-zero trace.  Solving the equations of motion one can easily show that this theory is gauge equivalent to the sigma model description of eq.~\eqref{eq:HQKsigma}.

Instead of sticking with the original $SU(n)_L\times SU(n)_R$ symmetry, consider taking the linear combination $SU(n)\times SU(n)^\prime$ defined in table \ref{tab:HQxi}.  The associated vector superfields are
\begin{align}
V & =\frac{1}{2}\nbrack{V_L+V_R} &
V^\prime & =\frac{1}{2}\nbrack{V_L-V_R}.
\end{align}
The gauge symmetry associated with $V$ is the anomaly free, diagonal combination with $\hat{h}_{R,n}=\hat{h}_{L,n}=\hat{h}_n$.  In terms of these new vector superfields the K\"ahler potential is
\begin{align}\label{eq:HQK}
K= & \Tr{(1-a)v^2\ln{(\xi_\eta^\dag\xi_\eta)}+(1-\tilde{a})\tilde{v}^2\ln{(\tilde{\xi}_\eta\tilde{\xi}_\eta^\dag)}}+ \nonumber\\
& \Tr{av^2\nbrack{\pfrac{\xi_\eta^\dag\xi_\eta}{d_\eta^\dag d_\eta}e^{V_{B^\prime}-V-V^\prime}-V_{B^\prime}}+\tilde{a}\tilde{v}^2\nbrack{\pfrac{\tilde{\xi}_\eta\tilde{\xi}_\eta^\dag}{d_\eta^\dag d_\eta}e^{V-V_{B^\prime}-V^\prime}+V_{B^\prime}}}.
\end{align}
All vector superfield equations of motion are solved for
\begin{align}\label{eq:HQVsol}
\xi_\eta^\dag\xi_\eta & = e^{2\bar{\kappa}_R}e^{V-V_{B^\prime}+V^\prime} &
\tilde{\xi}_\eta\tilde{\xi}_\eta^\dag & =e^{2\bar{\kappa}_R}e^{V_{B^\prime}-V+V^\prime}
\end{align}
after substituting $e^{\kappa_R}$ for $d$ and defining $2\bar{\kappa}_R=\kappa_R+\kappa_R^\dag$ as per section \ref{sec:RGS}.

At this point we chose to fix the gauge for $SU(n)^\prime$ and $U(1)_{B^\prime}$ by absorbing the vector superfields $V^\prime$ and $V_{B^\prime}$ into $\xi$ and $\tilde{\xi}$.  Conversely, we see that the chiral superfield $\kappa_R$, related to the spontaneously broken $R$-symmetry, could instead be considered as parameterising a gauge transformation.  Specifically, it would realise an HLS corresponding to the unbroken $U(1)_{R^\prime}$ symmetry, that has thus far been omitted from the HLS gauge group.  One can quite generally choose the Wess-Zumino gauge for subgroups of $\hat{H}$, then trade the scalar components of vector superfields for the corresponding quasi-NGBs (or vice-versa) in the above manner.

We therefore rewrite the vector superfields in the original K\"ahler potential as a sum of chiral superfields, then absorb them into dimensionful degrees of freedom
\begin{align}\label{eq:HQqdef}
V^\prime & =-\ln{(\sigma_n\sigma_n^\dag)} & V_{B^\prime} & =\ln{(\sigma_B\sigma_B^\dag)} &
q & =\frac{\sqrt{a}v\xi_\eta\sigma_n\sigma_B}{d_\eta} & \tilde{q} & =\frac{\sqrt{\tilde{a}}\tilde{v}\sigma_n\tilde{\xi}_\eta}{d_\eta\sigma_B}
\end{align}
where $\sigma_n$ transforms in the fundamental of $SU(n)^\prime$ such that $e^{-V^\prime}$ has the correct transformation properties.  Note that since $SU(n)^\prime$ is anomalous $V^\prime$ would only ever be expected to play the role of an auxiliary field in the low energy theory anyway.  Only $V$ can survive to become a true gauge field.

Eq.~\eqref{HQqxi} allows us to extract
$\sigma_n$ and $\sigma_B$ from $q$ and $\tilde{q}$ via the baryon expectations
\begin{align}\label{eq:HQsigmab}
\det{(\sigma_n)} & =\sqrt{\frac{\tilde{b}^{N+1\ldots N+n}b_{N+1\ldots N+n}}{(\sqrt{\tilde{a}a}\tilde{v}v)^n}} &
\sigma_B & =\sqrt{\frac{\sqrt{\tilde{a}}\tilde{v}(b_{N+1\ldots N+n})^{1/n}}{\sqrt{a}v(\tilde{b}^{N+1\ldots N+n})^{1/n}}}.
\end{align}
Hence the erstwhile gauge fields $V^\prime$ and $V_{B^\prime}$ can be replaced by $b$ and $\tilde{b}$.  That these expressions are given in terms of particular baryonic degrees of freedom is a side effect of our particular choice of electric quark expectation \eqref{eq:HQQvev}.  In full generality, one has
\begin{align}
\det{(\sigma_n\sigma_n^\dag)} & =\sqrt{\frac{(\tilde{b}\tilde{b}^\dag)(b^\dag b)}{(\tilde{a}a\tilde{v}^2v^2)^n}} &
\sigma_B\sigma_B^\dag & =\sqrt{\frac{\tilde{a}\tilde{v}^2(b^\dag b)^{1/n}}{av^2(\tilde{b}\tilde{b}^\dag)^{1/n}}}
\end{align}
and $q$ and $\tilde{q}$ are $SU(n)^\prime$ singlets transforming as
\begin{align}\label{eq:HQmgmat}
q & \in(\afund,\fund,{\bm1}) & \tilde{q} & \in(\fund,{\bm1},\afund)
\end{align}
under $SU(n)\times SU(N+n)_L\times SU(N+n)_R$.

Upon substituting in all new degrees of freedom we find the final form for the K\"ahler potential
\be\label{eq:HQKfin}
K=\Tr{q^\dag qe^{-V}+\tilde{q}\tilde{q}^\dag e^V}+v^2\ln{\pfrac{\det{(q^\dag q)}}{b^\dag b}}+\tilde{v}^2\ln{\pfrac{\det(\tilde{q}\tilde{q}^\dag)}{\tilde{b}\tilde{b}^\dag}}.
\ee
Explicit dependence on $a$ and $\tilde{a}$ is removed after eliminating $V_{B^\prime}$ using the vector superfield equations of motion.  The first terms here are simply the canonical K\"ahler potential of an $SU(n)$ gauge theory, under which $q$ and $\tilde{q}$ transform in the antifundamental and fundamental representations respectively.  This is precisely what we expect from the Seiberg dual, where the magnetic gauge field $V_{\rm mg}$ is identified with that of the diagonal $SU(n)$ gauge symmetry $V$.
All symmetry breaking is then driven by the remaining terms.

From eqs.~\eqref{HQqxi} and \eqref{eq:HQqdef} with all NGB expectations rotated to zero, the expectations of $q$ and $\tilde{q}$ are found to be
\begin{align}\label{eq:HQqvevs}
q & =\begin{pmatrix} 0 \\ b^{1/n}\uno \end{pmatrix} & \tilde{q} & =\begin{pmatrix} 0 & \tilde{b}^{1/n}\uno \end{pmatrix}.
\end{align}
An important observation is that the HLS description exhibits colour-flavour locking.  The $q$ and $\tilde{q}$ expectations break $SU(n)\times SU(N+n)_L\times SU(N+n)_R$ to $SU(N)_L\times SU(N)_R\times SU(n)_L\times SU(n)_R$, where $SU(n)_{L/R}$ is the diagonal combination of $SU(n)\subset SU(N+n)_{L/R}$ with the gauged $SU(n)$.  The orthogonal $SU(n)^\prime$ gauge symmetry, whose erstwhile gauge fields were absorbed into $q$ and $\tilde{q}$, does not mix with the flavour symmetry in this way.

\subsection{Magnetic mesons and the superpotential}

We have just shown how to derive the quark sector of the Seiberg dual theory, but the meson superfield and associated superpotential remain absent.  To find them, we consider the flipped coset sigma model description and use eq.~\eqref{eq:HQMdef} to define
\be\label{eq:HQMdef2}
M=v\tilde{v}e^{2\kappa_R}\begin{pmatrix} \tilde{\xi}_N\xi_N & \tilde{\xi}_N\xi_u \\ \tilde{\xi}_l\xi_N & \tilde{\xi}_l\xi_u \end{pmatrix}
\ee
which is a gauge singlet in the $(\afund,\fund)$ representation of $SU(N+n)_L\times SU(N+n)_R$, and is considered an elementary chiral superfield.  For simplicity we have chosen electric quark expectations with ${\bm v}=v\uno$ and $\tilde{\bm v}=\tilde{v}\uno$.\footnote{Otherwise the $v\tilde{v}$ prefactor is removed and one
makes the replacements $\xi_N\to{\bm v}\xi_N$ and $\tilde{\xi}_N\to\tilde{\xi}_N\tilde{\bm v}$, with similar generalisations in what follows.}

\begin{table}[!tb]
\be
\begin{array}{|c|c|cccc|}\hline
\widerow & SU(n) & SU(N+n)_L & SU(N+n)_R & U(1)_B & U(1)_R \\\hline
\widerow q & \afund & \fund & {\bm1} & \phantom{-}1/n\phantom{-} & N/(N+n) \\
\widerow \tilde{q} & \fund & {\bm1} & \afund & -1/n\phantom{-} & N/(N+n) \\
\widerow M & {\bm1} & \afund & \fund & 0 & 2n/(N+n) \\\hline
\end{array} \nonumber
\ee
\caption{The matter content of the full HLS description, which we identify with the usual magnetic theory.  The first $SU(n)$ is the gauge group.\label{tab:HQmg}}
\end{table}

Although it has been provocatively labelled with an ``$M$'' we should confirm that this object, derived from the flipped coset description, really can be interpreted as the superfield corresponding to $\tilde{Q}Q$.  To do so, we expand the quarks in the broken electric  theory around their expectations as
\begin{align}
Q & =\begin{pmatrix} v\uno+\delta Q & P \end{pmatrix} &
\tilde{Q} & =\begin{pmatrix} \tilde{v}\uno+\delta\tilde{Q} \\ \tilde{P} \end{pmatrix}
\end{align}
using components $\delta Q$ and $P$.  Normalised NGB superfields are then given explicitly by
\begin{align}
\Pi & =\frac{1}{v}\begin{pmatrix} \delta Q & P \\ 0 & 0 \end{pmatrix} &
\tilde{\Pi} & =\frac{1}{\tilde{v}}\begin{pmatrix} \delta\tilde{Q} & 0 \\ \tilde{P} & 0 \end{pmatrix}\, .
\end{align}
Using this basis to parameterise the Goldstone manifold we find from eq.~\eqref{eq:HLxiPi} that
\begin{align}\label{eq:HQxiQ}
\xi & =\begin{pmatrix} e^{\delta Q/v} & (e^{\delta Q/v}-\uno)\delta Q^{-1}P \\ 0 & \uno \end{pmatrix} &
\tilde{\xi} & =\begin{pmatrix} e^{\delta\tilde{Q}/\tilde{v}} & 0 \\ \tilde{P}\delta\tilde{Q}^{-1}(e^{\delta\tilde{Q}/\tilde{v}}-\uno) & \uno \end{pmatrix}.
\end{align}
Plugging into eq.~\eqref{eq:HQMdef2} and expanding to leading order gives
\be
M=\begin{pmatrix} v\tilde{v}\uno+v\delta\tilde{Q}+\tilde{v}\delta Q & \tilde{v}P \\ v\tilde{P} & \tilde{P}P \end{pmatrix}=\tilde{Q}Q
\ee
as required.  Note also that the meson expectation breaks the $SU(N)_L\times SU(N)_R$ factor of the flavour symmetry that is not broken by the expectations \eqref{eq:HQqvevs} of $q$ and $\tilde{q}$.

Having identified the meson superfield we define a real duality scale $\mu$ for normalisation, whereupon the superpotential
\be\label{eq:HQWdef}
W=\frac{1}{\mu}\Tr{Mq\tilde{q}}
\ee
is the unique choice compatible with eq.~\eqref{eq:HQmgmat}.  It is then straightforward to reconstruct the anomaly free $U(1)$ symmetries as in table \ref{tab:HQmg}.

Of course the superpotential is not merely allowed: it is required.  Simultaneously using normal and flipped coset descriptions means some of the degrees of freedom in $Q$ and $\tilde{Q}$ have been counted twice.  However, in conjunction with the expectation of $M$, the superpotential gives a mass $m_q=v\tilde{v}/\mu$ to $N$ flavours of $q$ and $\tilde{q}$.  Hence they are integrated out and the double counted degrees of freedom are removed.  Solving the equations of motion for the massive flavours and substituting back in, $W$ disappears and the sigma model description is again recovered.  Therefore the superpotential does not prevent the HLS and sigma model descriptions coinciding at low energy.

A second way of thinking about the superpotential is to recall that everything takes place on top of a non-zero meson background.  Ergo we can consider $M$ as as a source for $q\tilde{q}$.\footnote{We should point out that the superpotential should be considered as the 1PI effective superpotential for this interpretation to hold.  However, since the electric theory is completely higgsed it contains no massless, interacting degrees of freedom.  Hence the 1PI effective action successfully captures the low energy physics.  Equivalently, observe that the source in this case is nothing but a quark mass term so bestows the HLS description with a mass gap.}  It is then interesting to comment on the parallels between the superpotential and gauge sectors.  One can think of the superpotential as a sort-of chiral `gauge' coupling with the meson as its `gauge' field.

Just as the K\"ahler potential part of the theory deals with redundancies inherent to the NGB sector, so the superpotential deals with those in the quasi-NGB sector.  Indeed, the corresponding `gauge' transformations map $M\to g_RMg_L^\dag$, thereby moving one around the moduli space.  For appropriate supermultiplets (perhaps those of $\cN=2$ SUSY), it seems likely that the two sectors could be unified into a single current interaction. We will not explore this direction here.

Either way, integrating out the massive degrees of freedom leaves an SQCD-like theory with $n$ colours and $n$ flavours.  This is well known to confine \cite{Intriligator:1995au} due to its quantum deformed moduli space
\be\label{eq:HQQdefMS}
b^{N+1\ldots N+n}\tilde{b}_{N+1\ldots N+n}=m_q^N\Lambda_{\rm mg}^{2n-N}=(v\tilde{v})^N\mu^{-N}\Lambda_{\rm mg}^{2n-N}.
\ee
Here, $b$ and $\tilde{b}$ are the usual baryon degrees of freedom, we have set $q\tilde{q}=0$ and $\Lambda_{\rm mg}$ is the dynamical scale\footnote{By ``dynamical scale'' we are formally referring to the real scale at which the one loop RG equation for the corresponding gauge coupling diverges.} of the HLS description's gauge group.  We thus see that the superpotential drives confinement in the HLS description, and is therefore responsible for breaking the gauge symmetry.

So in addition to ensuring that the HLS description ultimately has the correct degrees of freedom, the superpotential also allows for a UV completion in which the gauge symmetry breaking is not simply encoded in the field definitions.  From the HLS point of view this confinement is inevitable because, once the magnetic quark degrees of freedom are made massive (as they must be in order to avoid double counting), it is the only way to reproduce the $G\rightarrow H$ symmetry breaking of the original nonlinear sigma model.

\subsection{Gauge symmetry restoration and the baryonic branch}

SQCD provides a perfect illustration of the discussion of symmetry restoration in section \ref{sec:RGS}.  To restore the gauge symmetry one takes the limit $e^{\bar{\kappa}_R}\to0$ holding $ve^{-\bar{\kappa}_R}$ and $\tilde{v}e^{-\bar{\kappa}_R}$ fixed.  It is the $U(1)_R$ symmetry and its associated quasi-NGB $\bar{\kappa}_R$
that enables magnetic quarks to be scaled in this way.  This is perhaps unsurprising, given that we know $R$-charges are generally related to scaling dimensions in superconformal field theories.

The decay constants, HLS gauge field masses and meson expectation \eqref{eq:HQMdef2} all vanish in this limit, with the magnetic quark masses following suit.  The gauge symmetry is therefore restored (as is the full flavour symmetry) and all additional terms vanish from eq.~\eqref{eq:HQKfin}, leaving $SU(n)$ SQCD+M with massless quarks and a canonical quark K\"ahler potential.

To recast the meaning of the limit in the language of Seiberg duality, recall that all order parameters arise from quasi-NGB expectations in the electric theory.  By choosing non-zero values for $v$ and $\tilde{v}$ we therefore define the duality at a particular point in moduli space, where the flavour symmetry is maximally broken.  Travelling away from this point in the magnetic theory, along the quasi-NGB direction that breaks $U(1)_R$, all fields (and consequently order parameters) are scaled to zero.

Duality implies that one should simultaneously move to the same point in the electric theory's moduli space, hence all order parameters vanish there as well.  In the full theory this would, of course, restore the corresponding electric gauge symmetry.  Note that the process is insensitive to the initial values of $v$ and $\tilde{v}$ so they remain independent parameters.

Before taking the
$e^{\bar{\kappa}_R}\to0$ limit the expectations of $Q$ and $\tilde{Q}$ saturated their ranks, forbidding expectations for the components of the electric quarks $P$ and $\tilde{P}$ (which instead contain NGBs of the symmetry breaking).  As the quark expectations start to vanish the constraint is relaxed and new flat directions open up; precisely those parameterised by the $\rho$-mesons constructed from $P$ and $\tilde{P}$.  At the same time the HLS gauge fields become massless.  Already we are starting to see hints that the magnetic gauge fields can be identified with electric $\rho$-mesons.

Note also that the magnetic quarks become massless as the meson expectation vanishes.  Initially, one may think this leads to double counted degrees of freedom in the HLS description.  However, the electric gauge fields become massless in this limit as well (besides which, the sigma model description breaks down so its degrees of freedom are ill defined).  Indeed, if we are to identify magnetic gauge fields with electric $\rho$-mesons, duality suggests we should also identify electric gauge fields with magnetic $\rho$-mesons.  Thus the newly massless magnetic quarks represent electric gauge fields rather than double counted degrees of freedom.

Another interesting limit to consider is the baryonic branch of the theory.  Here, one has a vanishing expectation for $\tilde{Q}$ but
\be
Q=\begin{pmatrix} v\uno & 0 \end{pmatrix}
\ee
i.e.\ only the baryon direction of moduli space has non-zero expectation.  The unbroken flavour symmetry is then
\be
H=SU(N)_L\times SU(n)_L\times SU(N+n)_R\times U(1)_{B^\prime}\times U(1)_{R^\prime}.
\ee
Hence all breaking is confined to the $SU(N+n)_L$ factor.

Following the standard procedure one does not expect magnetic antiquarks $\tilde{q}$ to appear in the HLS description at all.  The same is true for the meson superfield, for which only the components arising from $\xi_\eta^\prime$ are present.  Specifically, one would write down
\be\label{eq:HQKq}
K=\Tr{(1-a)v^2\ln{(\xi_\eta^\dag\xi_\eta)}+av^2\nbrack{\pfrac{\xi_\eta^\dag\xi_\eta}{d_\eta^\dag d_\eta}e^{V_{B^\prime}-V_L}-V_{B^\prime}}}
\ee
in place of eq.~\eqref{eq:HQK}.

The rest of the reasoning of section \ref{sec:RM} remains unchanged and yields a theory with
\be\label{eq:HQKq2}
K=\Tr{q^\dag qe^{-V}}+v^2\ln{\pfrac{\det{(q^\dag q)}}{b^\dag b}}
\ee
and no superpotential.  However, this theory clearly fails to capture all of the low energy physics as there are no degrees of freedom corresponding to the massless chiral superfields $\tilde{Q}$.  This is because no scalar components of $\tilde{Q}$ are NGBs on the baryonic branch so are not represented in a description based on NGB superfields.

Suppose we include the magnetic antiquarks, meson superfield
and $V_R$ anyway, and continue to use the K\"ahler potential \eqref{eq:HQKfin} with superpotential \eqref{eq:HQWdef}.  Expanding around the vacuum using eq.~\eqref{eq:HQqvevs} for the expectation of $q$, the superpotential becomes
\be
W=b^{1/n}(\tilde{Z}\delta\tilde{q}+Y\tilde{p})+\ldots\sep{where}
\tilde{q}=\begin{pmatrix} \delta\tilde{q} & \tilde{p} \end{pmatrix}\sep{and}
M=\mu\begin{pmatrix} X & \tilde{Z} \\ Z & Y \end{pmatrix}.
\ee
Hence all components of $\tilde{q}$ are massive.  Upon integrating them out the K\"ahler potential reduces back to eq.~\eqref{eq:HQKq} and the superpotential vanishes.  One can therefore include the magnetic antiquarks even though they are not explicitly generated by NGB superfields.  Once again the superpotential has proven itself instrumental in removing superfluous degrees of freedom from the HLS description.

Furthermore, the meson is now expanded to leading order as
\be
M=\tilde{Q}Q=v\begin{pmatrix} \delta\tilde{Q} & 0 \\ \tilde{P} & 0 \end{pmatrix}.
\ee
Its massless components $X$ and $Z$ are directly related to $\tilde{Q}$; degrees of freedom that would otherwise be missing.  Rather than arising from the flipped coset HLS description, the meson is now an extra matter superfield added by hand (or indeed just left alone for the whole procedure).  Meanwhile the superpotential ensures that the components $\tilde{Z}$ and $Y$, that correspond to degrees of freedom already accounted for in the quark sector, are massive and so removed from the low energy theory.

We therefore conclude that the HLS description derived in section \ref{sec:RM} also describes the baryonic branch of the theory, even though the symmetry breaking pattern is different there.  While the branch with $\tilde{v}=0$ is the only one discussed in this section, it is clear that the same reasoning applies to $v=0$.

\subsection{Summary}

In this section we have argued that the magnetic gauge group of Seiberg duality arises as an HLS, corresponding to the diagonal subgroup of the $SU(n)_L\times SU(n)_R$ flavour symmetry preserved at a generic point in moduli space.  The magnetic quarks and mesons are constructed from NGBs associated with the flavour symmetry breaking, with the superpotential $W=\mu^{-1}\Tr{Mq\tilde{q}}$ ensuring that no degrees of freedom are counted twice.  Equivalently, the combination of meson and superpotential can be thought of a source for $q\tilde{q}$ due to the fact that there is a non-zero meson background, or as a facilitator for a consistent UV completion of the HLS description.

The HLS formalism only defines the duality at points in moduli space where both electric and magnetic gauge groups are completely broken.  However, the presence of a spontaneously broken $R$-symmetry allows a limit to be taken in which both are restored.  The limit is smoothly attained by travelling along the quasi-NGB direction responsible for breaking $R$-symmetry until $R$-symmetry, and all other symmetries, are restored.

Although there are no explicit kinetic terms for the HLS gauge fields, we assume they are generated dynamically (as argued in ref.~\cite{Bando:1987br}) when the magnetic theory is in a higgsed phase, or when its gauge group is unbroken.  When it is in a confined phase we instead consider the kinetic terms to be suppressed by the divergent gauge coupling.  In either case, they are not forbidden by any symmetry so we are anyway free to add them by hand.\footnote{It is also plausible that kinetic terms are (at least partly) generated through a mixing with the a quasi-NGB $\bar{\kappa}_B$, as discussed at the end of section \ref{sec:RGS}.  To check for this, one could consider the full unbroken $U(1)_B$ generator when constructing $\xi$ and $\tilde{\xi}$ then factor out a chiral superfield $\kappa_B$, as was done with $\kappa_R$.  An $e^{\pm\bar{\kappa}_B}$ term would appear alongside the vector superfield terms in the K\"ahler potential, facilitating the mixing of the two.}

\section{Consequences of the HLS interpretation\label{sec:CH}}

Now that the HLS description of electric SQCD has been formally identified with the magnetic dual, we would like to use the interpretation to try and learn some more about Seiberg duality.  Our main results will be to confirm the identification of $\rho$-mesons with magnetic gauge fields (proposed in ref.~\cite{Komargodski:2010mc}), to improve our understanding of the effect of electric quark masses and to determine the value of the duality scale $\mu$ in certain regimes.

\subsection{$\rho$-mesons\label{sec:rho}}

The quickest way to understand the origin of the magnetic gauge fields is to consult the solution to their equations of motion \eqref{eq:HQVsol}.  These imply
\be
V_{\rm mg}=\frac{1}{2}\sbrack{\ln{(\eta\xi^\dag\xi\eta)}-\ln{(\eta\tilde{\xi}\tilde{\xi}^\dag\eta)}}.
\ee
At leading order in $1/v$ and $1/\tilde{v}$, eq.~\eqref{eq:HQxiQ} then gives
\be
V_{\rm mg}\approx\frac{1}{2}\sbrack{\ln{\nbrack{\uno+\frac{P^\dag P}{v^2}}}-\ln{\nbrack{\uno+\frac{\tilde{P}\tilde{P}^\dag}{\tilde{v}^2}}}}
\approx\frac{P^\dag P}{2v^2}-\frac{\tilde{P}\tilde{P}^\dag}{2\tilde{v}^2}
\ee
or, in terms of components
\be\label{eq:CHVmg}
V_{\rm mg}^\alpha\approx\Tr{S^\alpha\nbrack{\frac{P^\dag P}{v^2}-\frac{\tilde{P}\tilde{P}^\dag}{\tilde{v}^2}}}.
\ee
This unambiguously identifies the gauge fields of the magnetic theory with the $\rho$-mesons of the electric theory.

An orthogonal approach, using the colour-flavour locking observed in SQCD, is to examine the Noether currents attached to the unbroken $SU(n)_L$ and $SU(n)_R$ parts of the flavour symmetry.\footnote{This approach was used in ref.~\cite{Komargodski:2010mc} for a vacuum with broken SUSY\@.  The application here is practically identical.}  In electric variables they are
\begin{align}
\cJ^\alpha_L & =-\Tr{\vphantom{\tilde{P}^\dag}PS^\alpha P^\dag e^{V_{\rm el}}} &
\cJ^\alpha_R & =\Tr{\tilde{P}^\dag S^\alpha\tilde{P}e^{-V_{\rm el}}}
\end{align}
for an electric gauge superfield $V_{\rm el}$.  In the magnetic theory one has
\begin{align}
\cJ^\alpha_L & =\Tr{p^\dag S^\alpha pe^{-V_{\rm mg}}}+\mbox{mesons} &
\cJ^\alpha_R & =-\Tr{\tilde{p}S^\alpha\tilde{p}^\dag e^{V_{\rm mg}}}+\mbox{mesons}
\end{align}
where $q$ and $\tilde{q}$ have been parameterised using $n\times n$ matrices $p$ and $\tilde{p}$:
\begin{align}
q & =\begin{pmatrix} \delta q \\ p \end{pmatrix} & \tilde{q} & =\begin{pmatrix} \delta\tilde{q} & \tilde{p} \end{pmatrix}.
\end{align}

For small fluctuations in $V_{\rm el}$ and $V_{\rm mg}$ these currents can be expanded around the vacuum of the electric theory as
\begin{align}
\cJ^\alpha_L & \approx-\Tr{\vphantom{\tilde{P}^\dag}S^\alpha P^\dag P}+\mbox{3 particles} &
\cJ^\alpha_R & \approx\Tr{S^\alpha\tilde{P}\tilde{P}^\dag}+\mbox{3 particles}
\end{align}
or, expanding around magnetic expectations $p=b^{1/n}\uno$ and $\tilde{p}=\tilde{b}^{1/n}\uno$,
\begin{align}
\cJ^\alpha_L & \approx-\frac{1}{2}(b^\dag b)^{1/n}V_{\rm mg}^\alpha+\mbox{2 particles} &
\cJ^\alpha_R & \approx-\frac{1}{2}(\tilde{b}\tilde{b}^\dag)^{1/n}V_{\rm mg}^\alpha+\mbox{2 particles.}
\end{align}
Equating the two we can therefore write
\be\label{eq:CHVcfl}
V_{\rm mg}^\alpha\approx-\frac{1+c}{(b^\dag b)^{1/n}}\cJ_L^\alpha-\frac{1-c}{(\tilde{b}\tilde{b}^\dag)^{1/n}}\cJ_R^\alpha\approx\Tr{S^\alpha\nbrack{\frac{1+c}{(b^\dag b)^{1/n}}P^\dag P-\frac{1-c}{(\tilde{b}\tilde{b}^\dag)^{1/n}}\tilde{P}\tilde{P}^\dag}}
\ee
for any constant $c$.  This result is identical to that of eq.~\eqref{eq:CHVmg} for expectations $b^\dag b=(1+c)^nv^{2n}$ and $\tilde{b}\tilde{b}^\dag=(1-c)^n\tilde{v}^{2n}$.  Invoking the baryon map familiar from Seiberg duality we also have
\begin{align}
b^\dag b & =-(-\mu)^n\Lambda_{\rm el}^{n-2N}B^\dag B=-(-\mu)^n\Lambda_{\rm el}^{n-2N}v^{2N} \nonumber\\
\tilde{b}\tilde{b}^\dag & =-(-\mu)^n\Lambda_{\rm el}^{n-2N}\tilde{B}\tilde{B}^\dag=-(-\mu)^n\Lambda_{\rm el}^{n-2N}\tilde{v}^{2N}.
\end{align}
Hence $c$ is determined solely by the electric quark expectations:
\be\label{eq:CHcdef}
\pfrac{1+c}{1-c}^n=\pfrac{v}{\tilde{v}}^{2N-2n}.
\ee

All conclusions reached in this section are perturbative in nature and only apply when the fluctuations in $P$ and $\tilde{P}$ can be considered small.  These are generically of order $\Lambda_{\rm el}$, the dynamical scale of the electric theory, so one requires $v>\Lambda_{\rm el}$.  This limit will be emphasised when we consider the behaviour of electric and magnetic theories in section \ref{sec:mu}.

On a mesonic branch of the theory $\tilde{v}=v$ and we have $c=0$, which fixes $b=\tilde{b}=v^n$.  Hence the magnetic quarks are normalised so that their expectations, and therefore the symmetry breaking scale, correspond exactly to their electric counterparts (this idea will be useful when discussing the duality scale).  Referring back to eq.~\eqref{eq:HQqdef} and setting the expectations of
all fluctuations around the vacuum to zero then implies that $a=\tilde{a}=1$, as proposed for the mesonic branch in ref.~\cite{Kitano:2011zk}.

Baryonic branches have one of $v$ of $\tilde{v}$ equal to zero and so $c=\pm 1$.  Setting $\tilde{v}=0$, for example, implies that $c=1$, whereupon the magnetic gauge field is given by $V_{\rm mg}\approx2v^{-2n}\Tr{S^\alpha P^\dag P}$.  The absence of $\tilde{P}$ is is to be expected since the electric antiquarks no longer take part in the flavour symmetry breaking.  Furthermore, the magnetic quarks pick up a factor of $\sqrt{2}$ in their normalisation implying that $a=2$ is now the correct choice, as proposed for the baryonic branch in ref.~\cite{Komargodski:2010mc}.

\subsection{Electric quark masses\label{sec:EQM}}

Adding electric quark masses is well known to provide another way to higgs the magnetic gauge group.  The superpotential deformation
\be\label{eq:CHWel}
W_{\rm el}=-\Tr{{\bm m}\tilde{P}P}
\ee
for a rank $k\le n$ matrix ${\bm m}$ gives masses to $k$ flavours.  It also explicitly breaks the flavour symmetry to
\be
SU(N+n-k)_L\times SU(N+n-k)_R\times U(1)_B.
\ee
In the magnetic theory one adds the corresponding linear meson deformation
\be
\Delta W_{\rm mg}=-\mu\Tr{{\bm m}X}
\ee
for the $N\times N$ component of the meson $X$.  The F-terms for $X$ fix $\tilde{p}p=\mu{\bm m}$, higgsing the magnetic gauge group to $SU(n-k)$ at the origin of moduli space and giving mass to $k$ flavours of magnetic quark.  Duality is thus preserved at low energy, where massive flavours are integrated out of both descriptions.

From an HLS point of view we can understand this effect by varying the ratio $m/v$, where $m$ denotes a typical electric quark mass.  If $v\gg m$ the electric theory retains an approximate $SU(N+n)_L\times SU(N+n)_R$ flavour symmetry at the electric higgsing scale.  The HLS interpretation thus results in a magnetic theory based on a broken $SU(n)$ gauge group.

In line with our earlier discussion, the gauge group in this regime is broken by the confinement occurring once when the heavy magnetic quarks are integrated out.  The difference is a small correction to the gauge field masses from $\tilde{p}p\neq0$.  This only mildly breaks their degeneracy and is in accord with the approximate nature of the original $SU(N+n)_L\times SU(N+n)_R$ flavour symmetry.

If $v\ll m$ and $k<n$ the electric theory has $k$ fewer flavours at the electric higgsing scale.  This limits the possible symmetry breaking to
\be
H=SU(n-k)_L\times SU(n-k)_R\times U(1)_{B^\prime}.
\ee
Accordingly, the HLS description's gauge group is diminished to $SU(n-k)$.  Below the magnetic higgsing scale $\sqrt{\mu m}$ this matches the magnetic gauge group found through Seiberg duality.  Further still into the IR the magnetic theory again confines and breaks the residual gauge symmetry, as is usual in the HLS interpretation.

That the magnetic gauge group is completed to $SU(n)$ above the higgsing scale is necessary for continuity on the moduli space; as $v$ is increased we must recover the $m\ll v$ scenario.  Equivalently, the electric quark mass term lifts some of the quasi-NGBs parameterised by $\rho$-mesons.  Indeed, one expects $\rho$-mesons composed of quarks with mass $m>v$ to themselves have mass greater than $v$.  Due to their identification with these states, one expects the masses of the magnetic gauge fields to be similarly raised.

Of particular interest is the choice $k=n$.  Now the electric theory has $N$ colours and $N$ flavours below the scale $m$, so confines with its own quantum deformed moduli space
\be\label{eq:CHQdefel}
\det{\tilde{Q}Q}-\tilde{B}B=m^n\Lambda_{\rm el}^{2N-n}.
\ee
By adjusting $m/v$ we can then vary between higgsed and confined electric phases.  At the same time the magnetic theory varies between confined and higgsed phases.  We can exploit this effect to see that a {\em higgsed} $SU(n)$ HLS emerges from a {\em confining} $SU(N)$ gauge theory as we shall see in the following section.  In fact, the choice $v\ll m$ and $k=n$ can be recast as a reversal of the HLS interpretation.

\subsection{The duality scale\label{sec:mu}}

Another interesting corollary is the ability to fix the duality scale for a given electric quark expectation in certain regimes.  We generically find that $\mu$ should be chosen such that the scale of symmetry breaking is the same in both electric and magnetic theories.  In the following, we choose a specific, single point in moduli space with $\kappa_R=0$.

Consider first the case with no electric quark mass terms.  Using eq.~\eqref{eq:CHcdef} when $v=\tilde{v}$ we have $c=0$, ergo $b=\tilde{b}=v^n$ (up to phase factors).  Eq.~\eqref{eq:HQQdefMS} can then be solved for $\mu$:
\be
\mu^N=v^{2(N-n)}\Lambda_{\rm mg}^{2n-N}.
\ee
This should be compared with the relationship between electric and magnetic dynamical scales
\be
\mu^{N+n}=\Lambda_{\rm el}^{2N-n}\Lambda_{\rm mg}^{2n-N}
\ee
usually found in Seiberg duality.

Solving for $\mu$ we find
\be\label{eq:CHmu}
\mu=\Lambda_{\rm el}\pfrac{v}{\Lambda_{\rm el}}^{2(n-N)/n}=v\pfrac{v}{\Lambda_{\rm el}}^{(n-2N)/n}
\ee
and therefore
\be
\Lambda_{\rm mg}=\Lambda_{\rm el}\pfrac{v}{\Lambda_{\rm el}}^{2(N^2-n^2)/n(N-2n)}=v\pfrac{v}{\Lambda_{\rm el}}^{N(2N-n)/n(N-2n)}
\ee
for a given choice of $\Lambda_{\rm el}$ and $v$.  The magnitude of $\Lambda_{\rm mg}$ can also be compared to the magnetic quark mass
\be
m_q=\frac{v^2}{\mu}=\Lambda_{\rm el}\pfrac{v}{\Lambda_{\rm el}}^{2N/n}=v\pfrac{v}{\Lambda_{\rm el}}^{(2N-n)/n}
\ee
generated by the meson expectation, as well as the confinement scale of the magnetic theory $v$.

\begin{figure}
\begin{center}
\includegraphics[width=0.4\textwidth]{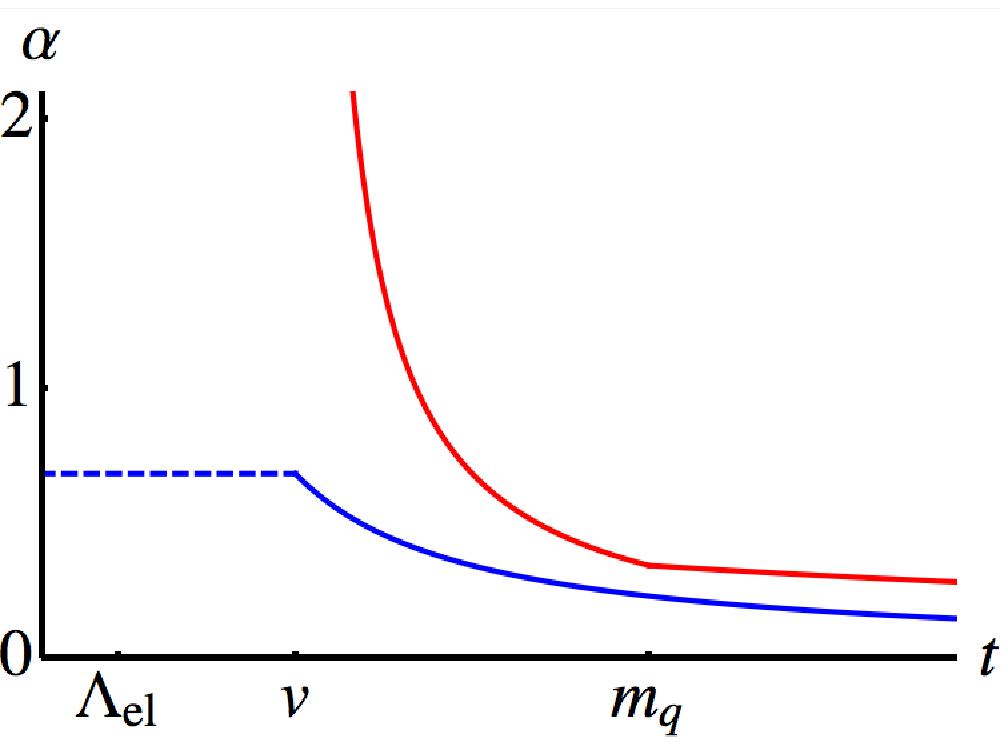}\hspace{1cm}
\includegraphics[width=0.4\textwidth]{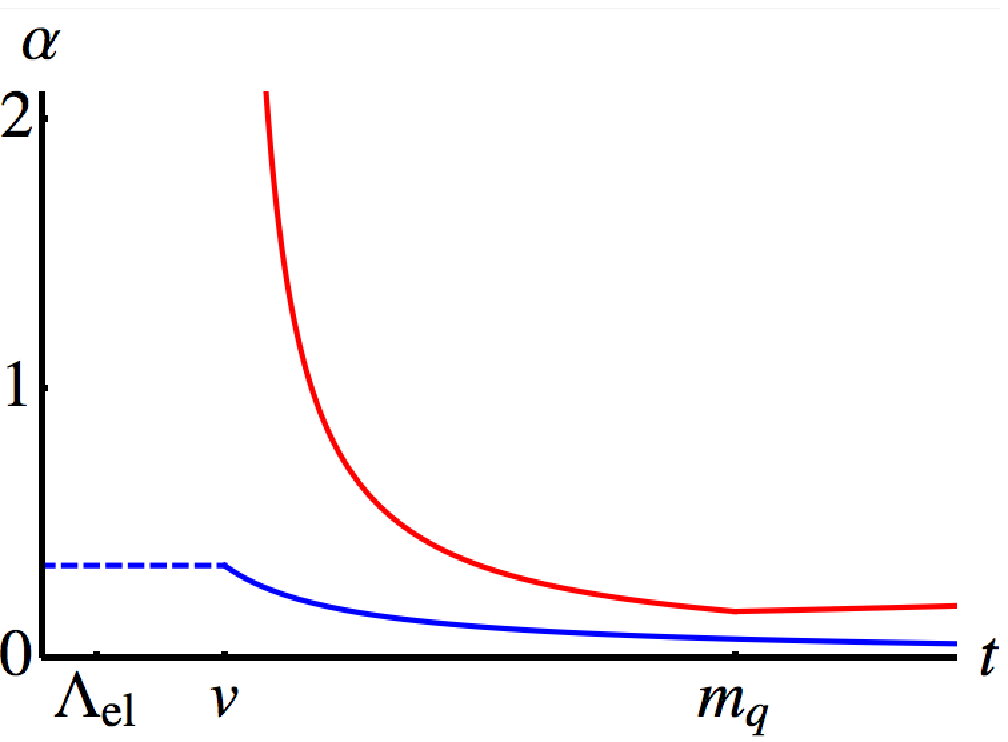}
\caption{One loop RG flows for the gauge coupling $\alpha=g^2/4\pi$ (as a function of logarithmic RG scale $t$) in massless SQCD when $v=10\Lambda_{\rm el}$.  Blue denotes the electric theory, red the magnetic and dashed lines a higgsed gauge group.  {\em Left}: the conformal window for $N=3$ and $n=2$.  {\em Right}: the free magnetic phase for $N=5$ and $n=2$.\label{fig:CHRGv}}
\end{center}
\end{figure}

These expressions are only valid when the perturbative arguments of section \ref{sec:rho} are valid, so can reliably be used to set $b=v^n$ and $\tilde{b}=\tilde{v}^n$; i.e.\ $v>\Lambda_{\rm el}$ such that fluctuations in $P$ and $\tilde{P}$ are small.  Choosing $N<2n<4N$ puts the theory in the conformal window.  Inspection of the above formulae then gives $\Lambda_{\rm mg}<v$.  Alternatively, choosing $N\ge2n$ puts us in the free magnetic phase and yields $\Lambda_{\rm mg}>v$.  In both cases the low energy behaviour is the same.  The magnetic theory confines at the scale $v$ whereas the electric theory is higgsed and provides a perturbative, low energy description.  Some example RG flows are shown in figure \ref{fig:CHRGv}.

Trying to take $v$ below $\Lambda_{\rm el}$ in this framework entails higgsing the electric theory in a strongly coupled electric regime, hence one looses some degree of theoretical control.  The magnetic theory also appears to confine below its dynamical scale in the conformal window, and above it in the free magnetic phase.  While the former is conceptually okay (despite not yielding a useful low energy description) the latter does not make physical sense.  This tells us that the assumptions used to fix $\mu$ break down, as expected.  It seems likely that, rather than being determined by eq.~\eqref{eq:CHmu}, $\mu$ is fixed at $\Lambda_{\rm el}$ for all choices $v<\Lambda_{\rm el}$.

Now consider turning on an electric quark mass \eqref{eq:CHWel} ${\bm m}=m\uno$, with $v=\tilde{v}$ hence $c=0$ maintained.  When $v>\sqrt{\mu m}$ there are no qualitative changes to the above; the higgsing scale of the electric theory remains higher than the confinement scale induced by the quark mass terms, i.e.\ $v>(m^n\Lambda_{\rm el}^{2N-n})^{1/2N}$.  Similarly, the confinement scale of the magnetic theory remains higher than the higgsing scale induced by the quark mass terms.  However, when $v=\sqrt{\mu m}$ (corresponding to $m=m_q$) the confinement and higgsing scales of electric and magnetic theories all become equal.

One cannot decrease $v$ any further than this; the electric quark expectations are fixed at $\sqrt{\mu m}$ by the geometry of the quantum deformed moduli space \eqref{eq:CHQdefel}.  Thus the scale appearing in eq.~\eqref{eq:CHVmg} should be replaced accordingly.  In addition, the magnetic quark expectations are no longer determined by confinement; they are fixed at $\sqrt{\mu m}$ by the higgsing superpotential.  Thus the scale appearing in eq.~\eqref{eq:CHVcfl} should also be replaced.

Upon updating both scales, the two expressions for the magnetic gauge field remain consistent provided $(\mu m)^N=m^n\Lambda_{\rm el}^{2N-n}$.  Equivalently, the magnetic quarks are still normalised such that magnetic and electric symmetry breaking scales are equal.  This is analogous to the result $b=\tilde{b}=v^n$ found in section \ref{sec:rho} on the mesonic branch, the implication being that the duality scale should always be chosen so as to enact this normalisation.

Consequently, one has
\begin{align}
\mu & =\Lambda_{\rm el}\pfrac{m}{\Lambda_{\rm el}}^{(n-N)/N}=m\pfrac{m}{\Lambda_{\rm el}}^{(n-2N)/N}
\end{align}
and therefore
\begin{align}
\Lambda_{\rm mg} & =\Lambda_{\rm el}\pfrac{m}{\Lambda_{\rm el}}^{(N^2-n^2)/N(N-2n)}=m\pfrac{m}{\Lambda_{\rm el}}^{n(2N-n)/N(N-2n)}
\end{align}
from the usual relationship between electric and magnetic scales.  Note that the parameter $v$ is now redundant.

\begin{figure}
\begin{center}
\includegraphics[width=0.4\textwidth]{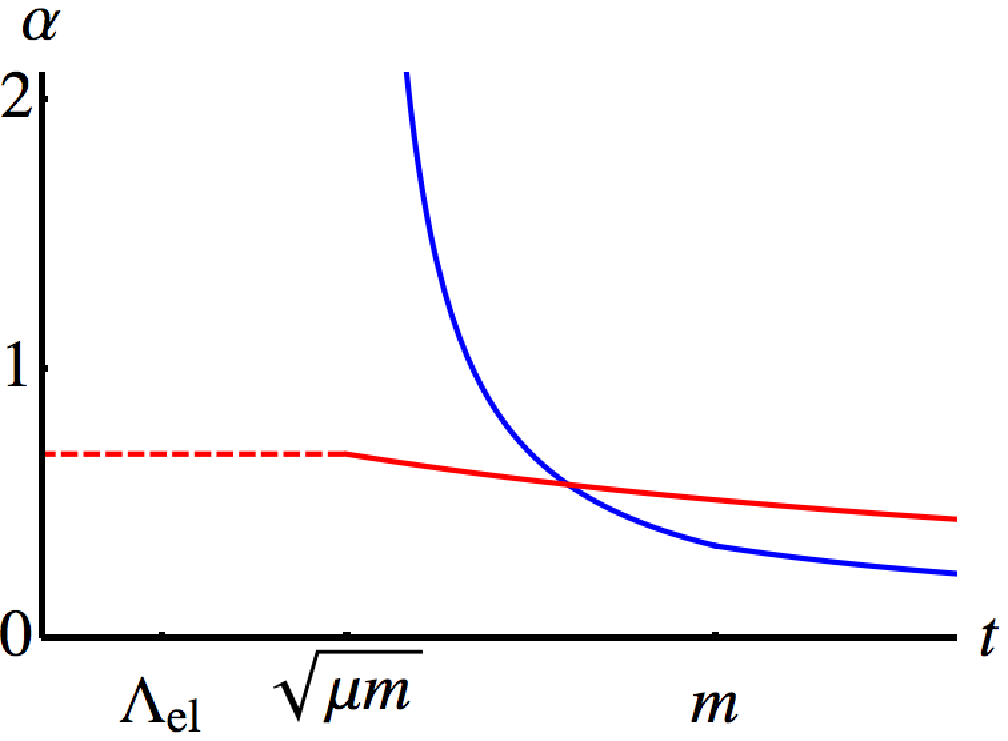}\hspace{1cm}
\includegraphics[width=0.4\textwidth]{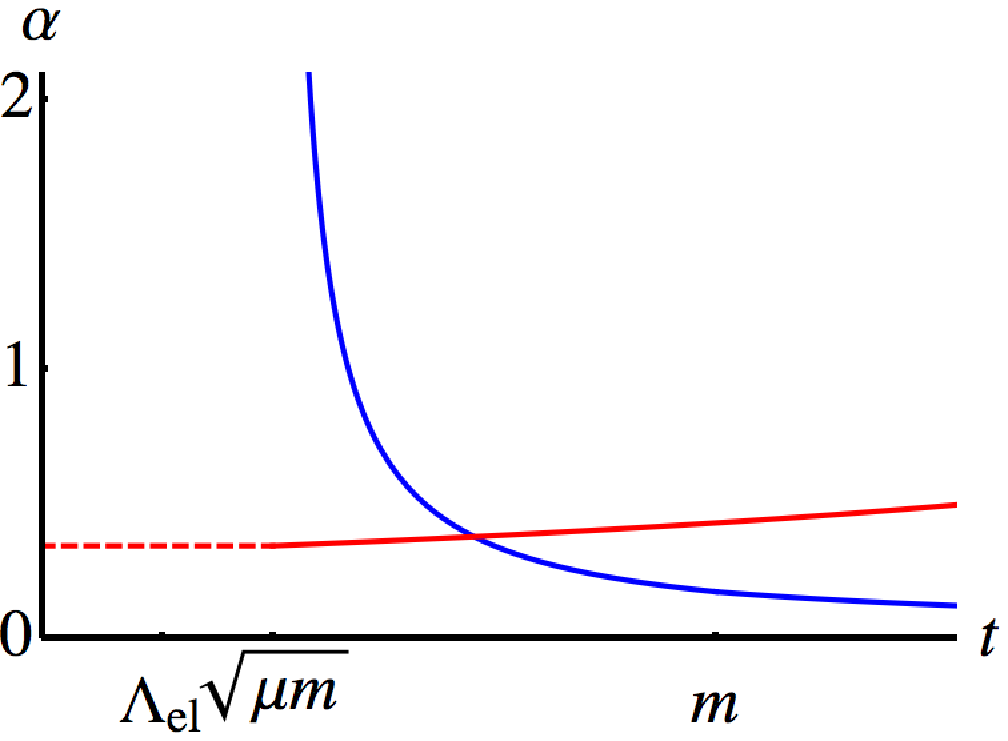}
\caption{One loop RG flows for the gauge coupling $\alpha=g^2/4\pi$ (as a function of logarithmic RG scale $t$) in massive SQCD when $v=0$ and $m=100\Lambda_{\rm el}>m_q$.  Blue denotes the electric theory, red the magnetic and dashed lines a higgsed gauge group.  {\em Left}: the conformal window for $N=3$ and $n=2$.  {\em Right}: the free magnetic phase for $N=5$ and $n=2$.\label{fig:CHRGm}}
\end{center}
\end{figure}

Choosing $m>\Lambda_{\rm el}$ one thus has $\Lambda_{\rm mg}<\sqrt{\mu m}$ in the conformal window and $\Lambda_{\rm mg}>\sqrt{\mu m}$ in the free magnetic phase.  The roles of electric and magnetic theories are then reversed: the electric theory confines in the IR, while the magnetic theory is higgsed and provides a perturbative, low energy description.  Some example RG flows are shown in figure \ref{fig:CHRGm}.

Just like $v$ in the massless case, trying to take $m$ below $\Lambda_{\rm el}$ leads to ill defined low energy physics in this framework.  This is because the expansion scale in eq.~\eqref{eq:CHVmg} is now $\sqrt{\mu m}$.  When $m>\Lambda_{\rm el}$, fluctuations of $P$ and $\tilde{P}$ are therefore sufficiently small for the expansion to be trusted.  At smaller values of $m$ the perturbative approximation is invalidated.

Variation between confined and higgsed phases in the magnetic theory is clear.  The most useful regime is the clearly latter, where the electric theory confines and can be described at low energy by a higgsed magnetic theory.  However, Seiberg duality is (of course) a duality, so the whole HLS interpretation should work both ways.  We will shortly see that this is indeed true, but a brief examination of figures \ref{fig:CHRGv} and \ref{fig:CHRGm} already reveals an electric/magnetic exchange symmetry with $v\leftrightarrow\sqrt{\mu m}$ and $m\leftrightarrow m_q$.

\section{Gauged $U(1)_R$\label{sec:GR}}

So far we have been focussing on how the HLS maps onto the magnetic description in Seiberg duality.  The HLS theory describes the same low energy physics as the $G^c/\hat{H}$ nonlinear sigma model of broken $SU(N)$ SQCD, which can be thought of as the linear UV completion.  However, it is interesting to ask what, given the magnetic HLS theory,  one can learn about the full $SU(N)$ electric theory if its form is not assumed a priori.

Consider the K\"ahler potential of the electric theory in the symmetry restoring limit, $K\supset Q^\dagger Q+\tilde{Q}^\dagger\tilde{Q}$, where we will take as read appropriate factors of $e^V$ to make it gauge invariant.  This term is proportional to the anomalous current $U(1)_A$, which is in turn related to the global $R$-current supermultiplet (as described in \cite{Abel:2011wv}).  Indeed, as we have seen, the modulus $\kappa_R$ for the breaking of $U(1)_R$ can play a role in the restoration of symmetry.  It can also be recast as a gauge field.  Therefore in order to answer this question one is motivated to first consider gauging $U(1)_R$.

This has to be carried out within the framework of superconformal ${\cal N}=1$ supergravity \cite{Ferrara:1983dh, Barbieri:1982ac, Chamseddine:1995gb}. The additional fields of interest for our discussion are the $U(1)_R$ gauge field $V_R$ and the conformal compensator $\phi$.  Assuming the usual $R$-charge for the gauginos of $+1$, the fields transform under a $U(1)_R$ transformation as 
\begin{align}
\phi & \longrightarrow e^{2i\chi/3}\phi \nonumber\\
V_R & \longrightarrow V_R+i(\chi-\chi^\dag) \nonumber\\
\phi_i & \longrightarrow e^{-iR_i\chi}\phi_i  
\end{align}
where $\phi_i$ are generic matter superfields (i.e.\ quarks and mesons) with superconformal $R$-charges $R_i$ and $\chi$ is a chiral superfield. Note that the extra scaling symmetry introduced by superconformality changes $\kappa_R$ from an M-type to a P-type superfield.  The general Lagrangian is of the form 
\begin{align}
\cL=\dimint{4}{\theta}\nbrack{\phi^\dag e^{-2V_R/3}\phi}K\nbrack{(e^V\phi)^i,\phi_i^\dag e^{R_iV_R }} & +\dimint{2}{\theta}\phi^3 W(\phi_i)+\mbox{h.c.}+ \nonumber\\
&+\mbox{gauge/gravity terms.}
\end{align}

For the superpotential $W$ to give an invariant term it should transform as $W\to e^{-i2\chi}W$, i.e.\ it has $R$-charge $+2$ as usual.\footnote{At the quantum level there also appear wavefunction renormalisation factors ${\cal Z}_i$.  These will be set to one here (they are discussed in ref.~\cite{Luty:1999qc}).  Also note this approach is not quite that taken in ref.~\cite{Luty:1999qc} -- the $V_R$ here is the actual $R$-gauge field rather than the gauge field of a normal anomaly free gauge group.}  Thus the relevant terms in the magnetic theory take the form
\begin{align}\label{rabbit}
\cL\supset\dimint{4}{\theta}\nbrack{\phi^\dag e^{-2V_R/3}\phi} & \Tr{q^\dag e^{R_qV_R}q+\tilde{q}^\dag e^{R_{\tilde{q}}V_R}\tilde{q}+\Phi^\dag e^{R_MV_R}\Phi+v^2\uno}+ \nonumber\\
\dimint{2}{\theta}\phi^3 & \Tr{\Phi q\tilde{q}}+\mbox{h.c.}
\end{align}
where we have defined the canonically normalised meson by $\Phi \sim M/ \mu $ and the trace is over the $n+N$ flavour indices (with contraction over $SU(n)$ colour being understood).  We have taken degenerate expectations in the FI-term.  The $R$-charges $R_q=R_{\tilde{q}}=N/(N+n)$ are fixed by absence of $SU(n)-SU(n)-R$ anomalies, while $R_M=2n/(N+n)$ is fixed by the superpotential coupling in eq.~\eqref{rabbit}: they are as given in table~\ref{tab:HQxi}.

As per the Seiberg magnetic dual this theory has an \emph{anomalous} global $U(1)_A$ symmetry, under which the quarks and antiquarks have charge $+1$ and the meson has charge $-2$.  In the normal HLS fashion one solves the equation of motion of the $V_R$ field
\be\label{zebra} 
\nbrack{R_q-\frac{2}{3}}q^\dag qe^{R_qV_R}+\nbrack{R_{\tilde{q}}-\frac{2}{3}}\tilde{q}^\dag\tilde{q}e^{R_{\tilde{q}}V_R}+\nbrack{R_\Phi-\frac{2}{3}}\Phi^\dag\Phi e^{R_M V_R}=\frac{2}{3}v^2\uno 
\ee
which gives 
\be
(q^\dag q+\tilde{q}^\dag\tilde{q})e^{R_qV_R}-2\Phi^\dag\Phi e^{R_M V_R}=\frac{3(N+n)}{N+2n}v^2\uno.
\ee
The LHS of this equation is the current $\cJ_A$, hence we can rewrite the equation as
\be\label{aardvark}
\cJ_A=\frac{3(N+n)}{N+2n}v^2\uno.
\ee

As we have seen, the magnetic theory spontaneously breaks the $SU(N+n)_L\times SU(N+n)_R$ flavour symmetry to the colour-flavour locked $H\simeq SU(n)\times SU(n)^\prime$ because $q^\dag q$ and $\tilde{q}^\dag\tilde{q}$ have rank $n$.  Thus the quarks contribute only rank $n$ to the LHS of eq.~\eqref{aardvark}, while the colour singlet mesons contribute the remaining rank $N$ expectation.  This breaks the original symmetry as $G\times H_{\rm(local)}\to H_{\rm(global)}$.  However, the RHS of eq.~\eqref{aardvark} is flavour symmetric and so, therefore, is the LHS\@.  Indeed the expectations (setting elements $q=\tilde{q}$) have to satisfy
\begin{align}\label{bumblebee}
q=\tilde{q} & \sim e^{-NV_R/2(N+n)} & \Phi & \sim e^{-nV_R/(N+n)}.
\end{align}
We shall return to this result below. 

Now suppose that we knew only this magnetic theory, and wanted to infer the properties of the electric one by mapping the equations above.  More precisely, let us assume that there are no colour singlet states in the UV completion, and determine the form that eq.~\eqref{aardvark} must take in terms of electric degrees of freedom.  By definition there is no $SU(n)$ HLS in the electric theory to mix with the original $SU(n)_L\times SU(n)_R$ flavour symmetry.  Hence a necessary requirement for the $H$ subgroup to remain unbroken is that the LHS of eq.~\eqref{aardvark} maps in the electric theory to the sum of two Hermitian $(N+n)\times(N+n)$ matrices, whose ranks are given by $[{\rm rank}(G)-{\rm rank}(H)]/2=N$ (one to break $SU(N+n)_L\rightarrow SU(n)_L$ and one to break $SU(N+n)_R\rightarrow SU(n)_R$).

It is a theorem that every rank $N$, Hermitian, $(N+n)\times(N+n)$ matrix $A$ has a rank factorisation $A=Q^\dag Q$, where $Q$ is an $N\times(N+n)$ matrix.  Therefore even when the electric theory is not weakly coupled, the LHS of eq.~\eqref{aardvark} is proportional to 
\be
{Q^\dag Q+\tilde{Q}\tilde{Q}^\dag}
\ee
where $Q$ and $\tilde{Q}^\dag$ are some $N\times(N+n)$ matrices.  For a weakly coupled UV completion, $Q$ and $\tilde{Q}$ are of course identified with the electric quarks transforming as fundamental and anti-fundamental under $SU(N)$.

Eq.~\eqref{zebra} written in terms of weakly coupled canonically normalised electric degrees of freedom must take the form 
\be\label{coyote} 
\nbrack{R_Q-\frac{2}{3}}Q^\dag Qe^{R_QV_R}+\nbrack{R_{\tilde{Q}}-\frac{2}{3}}\tilde{Q}^\dag\tilde{Q}e^{R_{\tilde{Q}}V_R}=\frac{2}{3}v^2 \uno.
\ee
Equating the LHS of eqs.\eqref{zebra} with the LHS of \eqref{coyote} we find consistency with the mapping of non-holomorphic operators discussed in \cite{Luty:1999qc,Abel:2011wv} which, among other things, enables one to track soft terms through strong coupling.

Absence of $SU(N)-SU(N)-R$ anomalies then fixes $R_Q=R_{\tilde{Q}}=n/(n+N)$, with the result that $Q\tilde{Q}\sim e^{-nV_R/(N+n)}\sim\Phi $.  Furthermore the other expectations in eq.~\eqref{bumblebee} are consistent with the mapping $q^n\to Q^N $ and with the classical constraints $\det{(M)}=B\tilde{B}$.  Of course, we knew from Seiberg duality that this had to be the case since these identifications are known to be consistent with $R$-symmetry.  However, since a field's $R$-charge directly determines its expectation when one solves the $V_R$ equation of motion, the role of $R$-symmetry in the matching of moduli spaces is self-evident.

Finally, if we begin with no meson in the magnetic theory (and no superpotential there), there is clearly a role reversal: the rank condition on the magnetic quarks leaves an $SU(N)_L\times SU(N)_R$ flavour symmetry unbroken, with the $SU(N)$ electric gauge symmetry acting as the HLS for the anomaly free diagonal subgroup.

\section{Variations of Seiberg duality\label{sec:VS}}

Seiberg duality also exists for gauge theories based on $SO$ and $Sp$ groups, so the HLS interpretation ought to work for these as well.  To show this, it is convenient to first invert the HLS interpretation so as to derive the electric description instead.  Dual gauge groups for $SO$ and $Sp$ theories are then determined by the rank of meson expectation that causes the magnetic theory to confine.  In some senses this is actually a more natural way to think of the HLS interpretation.  There is no need to introduce a superpotential and one automatically finds a confining original theory being described by a higgsed HLS theory.

Consider starting from a magnetic $SU$ theory with $n$ colours and $N+n$ flavours, then giving an expectation to $M$ to provide a mass term for some of the magnetic quarks.  The maximum rank for this expectation is $N$.  Larger values lead to an effective theory with fewer flavours than colours, in which case an ADS superpotential \cite{Affleck:1983mk,Affleck:1984xz} is generated and the theory no longer has a vacuum.

When the rank condition is saturated the maximum possible flavour symmetry is preserved by choosing
\be
M=v\begin{pmatrix} \uno & 0 \\ 0 & 0 \end{pmatrix}
\ee
whereupon $SU(N+n)_L\times SU(N+n)_R\times U(1)_B\times U(1)_R$ is broken to $SU(N)\times SU(n)_L\times SU(n)_R\times U(1)_B$.  Integrating out the $N$ massive flavours the theory confines (as described by eq.~\eqref{eq:HQQdefMS}), further breaking the flavour symmetry to $SU(N)\times SU(n)_L\times SU(n)_R$ and completely breaking the gauge symmetry.  Both surviving $SU(n)$ factors can be thought of as a mixture of flavour and global gauge transformations.  Note that it is vital that all gauge fields are rendered massive if we are to find a sigma model (and therefore HLS) description at low energy.  This will be important when we move onto the other gauge groups.

Splitting generators into those that are broken and those that are not we find broken generators
\begin{align}
\hat{T}_L & =\begin{pmatrix} \hat{T}_N+n\uno & \hat{T}_u \\ 0 & -N\uno \end{pmatrix} &
\hat{T}_R & =\begin{pmatrix} -\hat{T}_N-n\uno & 0 \\ \hat{T}_l & N\uno \end{pmatrix}
\end{align}
for a traceless, anti-Hermitian matrix $\hat{T}_N$ and arbitrary complex matrices $\hat{T}_u$ and $\hat{T}_l$.  There are also identity matrices for the two broken $U(1)$ symmetries.  The unbroken generators are
\begin{align}
\begin{pmatrix} \hat{S}_N & 0 \\ 0 & 0 \end{pmatrix} &&
\begin{pmatrix} 0 & 0 \\ 0 & \hat{S}_{L,n} \end{pmatrix} &&
\begin{pmatrix} 0 & 0 \\ 0 & \hat{S}_{R,n} \end{pmatrix}
\end{align}
for traceless, Hermitian matrices $\hat{S}_N$, $\hat{S}_{L,n}$ and $\hat{S}_{R,n}$.  Therefore
\begin{align}
\xi & =e^{\kappa_R}\begin{pmatrix} e^{\kappa_B}\xi_N & \xi_u \\ 0 & e^{\kappa_B}\uno \end{pmatrix} &
\tilde{\xi} & =e^{\kappa_R}\begin{pmatrix} e^{-\kappa_B}\xi_N^{-1} & 0 \\ \xi_l & e^{-\kappa_B}\uno \end{pmatrix}
\end{align}
where $\det{(\xi_N)}=1$.  The chiral superfields $\kappa_R$ and $\kappa_B$ originate from the broken $R$- and baryon number symmetries respectively.  There are two projection operators:
\begin{align}
\eta & =\begin{pmatrix} \uno & 0 \\ 0 & 0 \end{pmatrix} & \eta^\prime & =\begin{pmatrix} 0 & 0 \\ 0 & \uno \end{pmatrix}.
\end{align}
Each commutes with $\hat{S}$ so we are free to work with left or right cosets for $\xi$ and $\tilde{\xi}$.

The $\eta^\prime$ projection of $\hat{S}$ would lead to an $SU(n)$ gauge theory, simply reproducing the magnetic theory we first thought of.  We therefore consider the $\eta$ projection instead, which projects $\hat{S}$ down to the generators of $SU(N)^c$.  We choose cosets for $\xi$ and $\tilde{\xi}$ such that
\begin{align}\label{eq:VStran}
\xi & \longrightarrow \hat{h}\xi g_L^\dag & \tilde{\xi} & \longrightarrow g_R\tilde{\xi}\hat{h}^{-1}
\end{align}
under nonlinear flavour transformations and define
\begin{align}
\xi_\eta & =\eta\xi=e^{\kappa_R}\begin{pmatrix} e^{\kappa_B}\xi_N & \xi_u \end{pmatrix} &
\tilde{\xi}_\eta & =\tilde{\xi}\eta=e^{\kappa_R}\begin{pmatrix} e^{-\kappa_B}\xi_N^{-1} \\ \xi_l \end{pmatrix}.
\end{align}
The $SU(n)_L\times SU(n)_R$ part of the unbroken flavour symmetry is realised linearly in $g_L$ and $g_R$.  This leads to a sigma model description with
\be
K_S=v^2\Tr{f\nbrack{(\xi_\eta\xi_\eta^\dag)(\tilde{\xi}_\eta^\dag\tilde{\xi}_\eta)}}
\ee
for a polynomial $f$, which is the most general K\"ahler potential invariant under eq.~\eqref{eq:VStran} (and includes terms of the form \eqref{eq:HQKsigma}).

For the HLS description we write down a trial K\"ahler potential
\be
K_V=v^2\Tr{\xi_\eta\xi_\eta^\dag e^V+\tilde{\xi}_\eta^\dag\tilde{\xi}_\eta e^{-V}}.
\ee
There is no FI-term as $V=V^\alpha\hat{S}_N^\alpha$ is traceless.  Solving the equations of motion for $V$ gives
\be
\xi_\eta\xi_\eta^\dag e^V=\tilde{\xi}_\eta^\dag\tilde{\xi}_\eta e^{-V}
\ee
and so
\be
K_V=2v^2\Tr{\sqrt{(\xi_\eta\xi_\eta^\dag)(\tilde{\xi}_\eta^\dag\tilde{\xi}_\eta)}}.
\ee
To reproduce the sigma model description we therefore write down a full HLS description K\"ahler potential
\be
K=v^2\Tr{f\nbrack{(\xi_\eta\xi_\eta^\dag)(\tilde{\xi}_\eta^\dag\tilde{\xi}_\eta)}}+av^2\Tr{\xi_\eta\xi_\eta^\dag e^V+\tilde{\xi}_\eta^\dag\tilde{\xi}_\eta e^{-V}-2\sqrt{(\xi_\eta\xi_\eta^\dag)(\tilde{\xi}_\eta^\dag\tilde{\xi}_\eta)}}.
\ee

Defining dimensionful chiral superfields $Q=\sqrt{a}v\xi$ and $\tilde{Q}=\sqrt{a}v\tilde{\xi}$, and absorbing all non-gauge terms in $f$ this becomes
\be
K=\Tr{QQ^\dag e^V+\tilde{Q}^\dag\tilde{Q}e^{-V}}+v^2\Tr{f\nbrack{\frac{(QQ^\dag)(\tilde{Q}^\dag\tilde{Q})}{v^4}}}
\ee
i.e.\ an $SU(N)$ gauge theory with $N+n$ flavours of electric quark $Q$ and $\tilde{Q}$.  The definitions of $Q$ and $\tilde{Q}$, along with the constraint $\det{(\xi_N)}=1$, determine their expectations to be rank $N$ and of order $\sqrt{a}v$.  The electric gauge group is thus completely higgsed, as required.

\subsection{$SO$ dualities}

Seiberg duality for $SO$ gauge groups \cite{Seiberg:1994pq,Intriligator:1995au,Intriligator:1995id} can be summarised in table \ref{tab:VSSO}.  Starting from a magnetic theory with $n+4$ colours (i.e.\ gauge group $SO(n+4)$) and $N+n$ flavours, we again consider giving an expectation to the magnetic meson.  Because of the superpotential this still give masses to magnetic quarks.

\begin{table}[!tb]
\be
\begin{array}{|c|c|cc|}\hline
\widerow & SO(n+4) & SU(N+n) & U(1)_R \\\hline
\widerow q & \fund & \afund & (N-2)/(N+n) \\
\widerow M & {\bm1} & \sym & 2(n+2)/(N+n) \\\hline\hline
\widerow & SO(N) & SU(N+n) & U(1)_R \\\hline
\widerow Q & \fund & \fund & (n+2)/(N+n) \\\hline
\end{array} \nonumber
\ee
\caption{The matter content of the magnetic (top) and electric (bottom) theories for $SO$ gauge groups.  The magnetic theory has superpotential $W=(1/\mu)\Tr{Mqq}$.\label{tab:VSSO}}
\end{table}

The maximum rank allowed for the meson expectation is $N$, above which the theory has no vacuum due to the generation of an ADS superpotential.  Assuming the rank condition is saturated, the flavour symmetry is broken to $SO(N)\times SU(n)$ as the meson is in the symmetric representation of $SU(N+n)$.  At the same time the gauge sector of the magnetic theory confines after integrating out the massive quarks, so a low energy sigma model description is appropriate.

Since the generators of $SO(n+4)$ are antisymmetric tensors one can define the hybrid baryon
\be
b^{\prime\prime}_{j_1\ldots j_n}=\epsilon_{\alpha_1\ldots\alpha_{n+4}}{\cal W}_\alpha^{\alpha_1\alpha_2}{\cal W}_\alpha^{\alpha_3\alpha_4}q^{\alpha_5}_{j_1}\ldots q^{\alpha_{n+4}}_{j_n}
\ee
where ${\cal W}_\alpha$ (being careful not to confuse the spinor index $\alpha$ with the gauge indices $\alpha_i$) is the antisymmetric field strength superfield.  The F-terms for $M$ fix the expectation of $q$ to be at most rank $n$, so this is the only baryon-like operator whose expectation does not vanish.  It is a singlet under the residual $SU(n)$ so a non-zero expectation for $b^{\prime\prime}$ leaves this part of the flavour symmetry intact.  At the quark level, this can be understood as a mixing between real $SU(n)$ flavour transformations and complexified, global $SO(n)\subset SO(n+4)$ gauge transformations. 

We then proceed exactly as before to find an HLS description with gauge group $SO(N)$, where the unbroken $SU(n)$ part of the flavour symmetry is realised linearly.  Since there is only one $SU(N+n)$ factor in the flavour symmetry there are no $\tilde{Q}$ superfields and the K\"ahler potential is
\be\label{eq:VSK}
K=v^2\Tr{\ln{(QQ^\dag)}}+\Tr{QQ^\dag e^V}
\ee
for $SO(N)$ gauge field $V$, and $Q$ transforming as in table \ref{tab:VSSO}.  This is exactly the electric theory anticipated.  Note that the associated metric is smooth everywhere so there is no problem in taking the symmetry restoring limit $Q\to0$.

Furthermore, both the gauge and $SU(N)$ part of the flavour symmetry are broken by a maximal, rank $N$ expectation of $Q$ following from the usual definition of $Q$.  This in turn implies a rank $N$ expectation for the electric meson $QQ$ and baryon $B=Q^N$.  The mapping of these expectations to their magnetic counterparts
\begin{align}
M^{ij} & \longleftrightarrow Q^{\alpha i}Q^{\alpha j} &
b^{\prime\prime}_{j_1\ldots j_n} & \longleftrightarrow\epsilon_{j_1\ldots j_nj_{1+n}\ldots j_{N+n}}B^{j_{1+n}\ldots j_{N+n}}
\end{align}
agrees with that expected from Seiberg duality.

\subsection{$Sp$ dualities}

$Sp$ theories are extremely similar to $SO$ theories from an HLS point of view.  Duality for them \cite{Intriligator:1995ne} is summarised in table \ref{tab:VSSp}.  Starting from a magnetic theory with $2n-4$ colours (i.e.\ gauge group $Sp(2n-4)$ -- the factor of 2 ensuring that the number of colours is even) and $2N+2n$ flavours, we take the familiar approach of giving an expectation to the magnetic meson.  Magnetic quarks are again rendered massive and, for this class of theory, the maximum allowed rank for the meson expectation is $2N+2$.  This leads to confinement with total flavour symmetry breaking, ergo no $H$ to form the basis of an HLS description.

Taking the next largest rank of $2N$ results in flavour symmetry breaking $SU(2N+2n)\to Sp(2N)\times SU(2n)$, as the meson is in the antisymmetric representation of $SU(2N+2n)$.  The theory goes on to confine after integrating out the massive quarks, but with no further flavour symmetry breaking.  The HLS description should therefore have gauge group $Sp(2N)$ and the $SU(2n)$ part of the unbroken symmetry is realised linearly.  Again, we explicitly point out that the confinement is vital in order for a sigma model description to apply at low energy.

The K\"ahler potential is that of eq.~\eqref{eq:VSK}, but with $V$ an $Sp(2N)$ gauge superfield, and $Q$ transforming as in table \ref{tab:VSSp}.  The conventional Seiberg dual electric theory is therefore recovered.  A rank $2N$ expectation for $Q$ breaks the gauge and $SU(2N)$ part of the flavour symmetry.  As for the $SO$ version there are no problems in the symmetry restoring limit $Q\to0$ and the mapping
\be
M^{ij}\longleftrightarrow Q^{\alpha i}Q^{\alpha j}
\ee
is in agreement with that predicted by Seiberg duality.

\begin{table}[!tb]
\be
\begin{array}{|c|c|cc|}\hline
\widerow & Sp(2n-4) & SU(2N+2n) & U(1)_R \\\hline
\widerow q & \fund & \afund & (N+1)/(N+n) \\
\widerow M & {\bm1} & \asym & 2(n-1)/(N+n) \\\hline\hline
\widerow & Sp(2N) & SU(2N+2n) & U(1)_R \\\hline
\widerow Q & \fund & \fund & (n-1)/(N+n) \\\hline
\end{array} \nonumber
\ee
\caption{The matter content of the magnetic (top) and electric (bottom) theories for $Sp$ gauge groups.\label{tab:VSSp}}
\end{table}

\subsection{Adjoint $SU(N)$ SQCD}

Theories with colour adjoints $X$ were examined in \cite{Kutasov:1995ve, Kutasov:1995np, Kutasov:1995ss}.  There is convincing evidence that in the presence of a superpotential for the adjoints
\be
W=\Tr{X^{k+1}}
\ee
an $SU(N)$ electric theory with $N_f$ flavours of quarks and anti-quarks maps to a magnetic $SU(n=kN_f-N)$ theory with its own adjoint $x$ and superpotential $W\supset\Tr{x^{k+1}}$.  The particle content in both theories is shown in table \ref{tab:KSSem}.

\begin{table}[!tb]
\begin{equation}\nonumber
\begin{array}{|c|c|c|c|c|c|}\hline
\widerow & SU(kN_f-N) & SU(N_f)_L & SU(N_f)_R & U(1)_B & U(1)_R \\\hline
\widerow q & \afund & \fund & \brm{1} & \phantom{-}\frac{1}{n}\phantom{-} & 1-\frac{2}{k+1}\frac{n}{N_f} \\\hline
\widerow \tilde{q} &\fund & \bm{1} &\afund & -\frac{1}{n}\phantom{-} & 1-\frac{2}{k+1}\frac{n}{N_f} \\\hline
\widerow x & \brm{adj} & \brm{1} & \brm{1} & 0 & \frac{2}{k+1} \\\hline
\widerow M_j & \brm{1} & \afund & \fund & 0 &  2-\frac{4}{k+1}\frac{N}{N_f} + \frac{2j}{k+1} \\\hline\hline
\widerow & SU(N) & SU(N_f)_L & SU(N_f)_R & U(1)_B & U(1)_R \\\hline
\widerow Q & \fund & \afund & \brm{1} & \phantom{-}\frac{1}{N}\phantom{-} & 1-\frac{2}{k+1}\frac{N}{N_f} \\\hline
\widerow \tilde{Q} &\afund & \bm{1} &\fund & -\frac{1}{N}\phantom{-} & 1-\frac{2}{k+1}\frac{N}{N_f} \\\hline
\widerow X & \brm{adj} & \brm{1} & \brm{1} & 0 & \frac{2}{k+1} \\\hline
\end{array}
\end{equation}
\caption{The content of the magnetic (top) and electric (bottom) theory in adjoint SQCD, where $j=0\ldots k-1$.\label{tab:KSSem}}
\end{table}

In order to understand this duality we first consider the moduli space of the electric theory.  Here the adjoint field enhances the moduli space.  Indeed the mesons are given by $M_j=\tilde{Q} X^j Q$, where $j=0\ldots k-1$.  The $X$ equation of motion ensures that the chiral ring is truncated as $X^{k}=0$.  Flavour can in principle be broken by giving expectations to ``dressed'' quarks $Q_j=X^j Q$.  Since these directions are not distinguished in the moduli space, their expectations break the enhanced complex flavour group $G^c=SU(kN_f)_L^c\times SU(kN_f)_R^c\times U(1)_B^c\times U(1)_R^c$.  At this point the rank condition in the electric theory still holds: we can only assign a rank $N$ expectation to the combined $Q_j$ and $\tilde{Q}_j$ system, and the flavour group is broken to $SU(kN_f-N)_L\times SU(kN_f-N)_R\times U(1)_{B^\prime}\times U(1)_{R^\prime}$.

Note that there are $2N_fN+N^2-1$ degrees of freedom available for the dressed quarks, and we require $2kN_fN$ degrees of freedom for the P and M-type superfields.  Hence we require $2N_f+N>2kN_f$ which we can write as
\be\label{banana}
N_f+N>(2k-1)N_f.
\ee
The free magnetic window of adjoint SQCD is $2N>(2k-1)N_f>(2k-1)(N+1)/k$, with $kN_f=N+1$ signalling $s$-confinement.  Since by assumption $N_f>N$, the condition in eq.~\eqref{banana} defining when we are able to use the HLS formalism also seems to define the upper edge of the free magnetic window.

The HLS description now closely follows the discussion in Section 3, but with the flavour symmetries enhanced to $SU(kN_f)$ factors, and with the HLS being assigned to the anomaly free diagonal $SU(kN_f-N)_L\times SU(kN_f-N)_R$ group.  In the standard coset description, $\xi_u$ and $\tilde{\xi}_l$ are $N\times(kN_f-N)$ matrices, with the latter index being identified with magnetic colour.  The full $\xi_\eta$ transforms as $\xi_\eta \rightarrow g_L\xi_\eta\hat{h}^{-1}_{L,kN_f-N}$ and similar for $\tilde{\xi}_\eta$, where $g_L$ are flavour rotations in $SU(kN_f)$, giving precisely the degrees of freedom of the dressed magnetic quarks.

The flipped coset description producing the mesons goes over all $SU(kN_f)$ indices and hence reproduces all $M_{j=0\ldots k-1}$ mesons.  The discussion extends quite directly to the more complicated $SU(N)$ duals discussed in refs.~\cite{Brodie:1996vx, Brodie:1996xm} -- for example the SQCD model presented there with two adjoints has $3kN_f$ flavours of dressed quarks, and the magnetic gauge group is accordingly $SU(3kN_f-N)$.

\section{Applications\label{sec:AP}}

Our main aim in this work has been to place Seiberg duality on a more dynamical footing.  In this way we hope ultimately to use the duality to learn more about dynamical processes in strongly coupled theories, rather than just the properties of the Lagrangians and vacua.  This section briefly summarises some of the applications we have in mind.  Detailed investigations are left for future work.

\subsection{Composite gauge fields}

An obvious application is composite gauge field scattering, as illustrated in figure \ref{fig:CHRGm}.  An electric theory where $n$ flavours have equal mass $m>\Lambda_{\rm el}$ is taken to be the `true' theory, but it confines in the IR meaning that the low energy physics is obscured.  Fortunately, Seiberg duality steps in and provides an alternative, perturbative description in the shape of a higgsed magnetic theory.  The magnetic gauge group is an emergent symmetry, with the massive gauge fields originating purely from composite operators.

If this were true, one would expect to see effects from the underlying electric theory near to the confinement scale.  In particular, signs of compositeness should start to appear in magnetic gauge field scattering amplitudes.  Using the HLS interpretation we can start to quantify such phenomena.  As discussed in sections \ref{sec:rho} and \ref{sec:mu}, magnetic gauge fields are explicitly related to electric $\rho$-mesons via
\be
V_{\rm mg}^\alpha\approx\frac{1}{\mu m}\Tr{S^\alpha(P^\dag P-\tilde{P}\tilde{P}^\dag)}
\ee
where
\be
\sqrt{\mu m}=\Lambda_{\rm el}\pfrac{m}{\Lambda_{\rm el}}^{n/2N}
\ee
is the higgsing scale of the magnetic theory, and also the confinement scale of the electric theory.

Schematically, one expects elastic scattering amplitudes for the longitudinal components of magnetic gauge bosons to grow with the centre of mass energy squared $s$.  This divergence is the standard unitarity violation problem.  It is addressed in the current framework at the higgsing scale of the magnetic theory, whereupon Higgs boson exchange unitarises the scattering.  The associated Higgs field is itself a composite object so, in effect, we have a composite Higgs model.

\begin{figure}
\begin{center}
\includegraphics[width=0.4\textwidth]{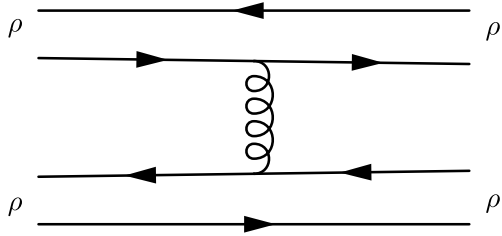}
\caption{The general form of the leading order processes contributing to $\rho$-meson scattering in a perturbative electric theory.\label{fig:PArhos}}
\end{center}
\end{figure}

However, the higgsing description is short lived and the composite nature of the gauge bosons immediately becomes apparent.  Their amplitudes are instead mapped onto the equivalent $\rho$-meson scattering amplitudes in the electric theory.  In the perturbative regime the leading order contribution of figure \ref{fig:PArhos} dominates, scaling as $\alpha(s)/s$ for running electric gauge coupling $\alpha(s)$.  Asymptotic freedom ensures that these amplitudes remain under control as the centre of mass energy is increased further.

\begin{figure}
\begin{center}
\includegraphics[width=0.4\textwidth]{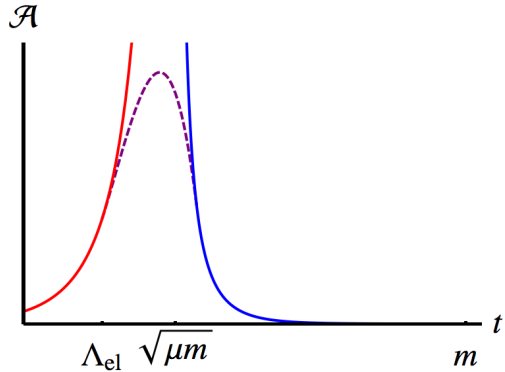}
\caption{The schematic behaviour of elastic scattering amplitudes (as a function of $t=\ln{s}$) for the longitudinal components of magnetic gauge bosons.  Below the higgsing scale the amplitude grows with $s$ before hitting a Higgs resonance at $\sqrt{\mu m}$.  Above the higgsing scale it matches onto the amplitude for elastic  $\rho$-meson scattering in the electric theory, which goes like $\alpha(s)/s$ in the perturbative regime. The dashed line crudely interpolates between the two perturbative regimes, tracing out a top heavy, broadened `resonance'.\label{fig:PACS}}
\end{center}
\end{figure}

The overall situation is illustrated in figure \ref{fig:PACS}.  Approaching from below the higgsing scale we anticipate the appearance of a resonance in $V_{\rm mg}V_{\rm mg}\to V_{\rm mg}V_{\rm mg}$ scattering, corresponding to the magnetic Higgs boson.  Approaching from above we expect the $\rho\rho\to\rho\rho$ amplitude to diverge as the electric theory becomes strongly coupled.  Interpolating between the two therefore suggests a top heavy, broadened `resonance' around the higgsing/confinement scale $\sqrt{\mu m}$.

This could, for example, be the case in the electroweak sector of the Standard Model.  All we really know is that low energy physics is well described by a broken $SU(2)\times U(1)$ gauge theory.  We do not know that this is ever an actual symmetry of nature.  It could merely emerge as an effective description of some other, strongly coupled theory.

One can also consider weakly gauging the original diagonal $SU(n)$ flavour factor.  In this case we denote the full gauge group as $SU(n)_c\times SU(n)_e$.  $SU(n)_c$ is the usual, composite, magnetic colour factor and $SU(n)_e$ is the elementary factor from the gauged flavour symmetry.  The only difference in the resulting HLS description is that the surviving diagonal subgroup
\be
SU(n)_c\times SU(n)_e\longrightarrow SU(n)
\ee
is now a gauge symmetry, which comes with {\em partially} composite gauge fields.

Taking gauge couplings $g_c$ and $g_e$ for each factor, the tree level mass eigenstates are
\begin{align}
V_h & =\frac{g_cV_c-g_eV_e}{\sqrt{g_c^2+g_e^2}} &
V_l & =\frac{g_eV_c+g_cV_e}{\sqrt{g_c^2+g_e^2}}
\end{align}
with mass squared $(g_c^2+g_e^2)\mu m$ and zero respectively.  The running gauge couplings are evaluated at the higgsing scale $\sqrt{\mu m}$ so the composition of the mass eigenstates can vary.  In the free magnetic phase, $g_c$ increases and the heavy state becomes more composite for larger values of $\sqrt{\mu m}$.  When $\sqrt{\mu m}=\Lambda_{\rm mg}$, the coupling $g_c$ hits its Landau pole and the heavy state is fully composite.

Models with partially composite gauge fields arising from Seiberg duality have recently been studied in the context of the electroweak sector of the MSSM \cite{Abel:2010vb, Craig:2011ev, Csaki:2011xn, Csaki:2012fh}.  It has been suggested that they have several phenomenological advantages, including an increased Higgs mass and a ``natural'' superpartner spectrum with light stops.\footnote{We should make the parenthetical remark that this second observation is based on the RG flow of the anomalous current operator, and it is therefore reliant on underlying assumptions about the initial pattern of SUSY breaking -- that it is universal, for example.  With a generic pattern of SUSY breaking mass squareds, the RG flow would simply expose those components that are proportional to anomaly free currents and which therefore map trivially (c.f.\ ref.~\cite{Abel:2011wv}). Since the latter are  traceless that would in turn imply very undesirable tachyonic mass squareds in the IR.}  In these models the symmetry breaking typically occurs in two phases, corresponding to two different electric quark masses $m_1>m_2$.  At the higher scale $\sqrt{\mu m_1}$ the $SU(2)_c\times SU(2)_e$ gauge symmetry is broken to its diagonal subgroup, which is identified with the $SU(2)$ of the MSSM\@.  At $\sqrt{\mu m_2}$ this is further broken by the usual (composite) Higgs fields.

For $m_1\gg m_2$ the two processes are well separated and the final $SU(2)$ breaking looks like a vanilla Higgs mechanism (albeit with a potential different from that usually found in the MSSM)\@.  For $m_1\sim m_2$ this separation does not exist and the underlying strong coupling has a large effect.  The final $SU(n)$ breaking starts to look more like that of a technicolour model, being driven by confinement in the electric theory.  Between these two regimes one has a composite Higgs model, with the compositeness becoming more noticeable as the ratio $m_1/m_2$ is increased.  Hence this framework can be used to continuously interpolate between higgsing and technicolour descriptions of (supersymmetric) electroweak symmetry breaking.

The HLS interpretation allows a concrete UV completion (i.e.\ electric theory) to be defined for these models.  It also allows for the exploration of phenomenology away from the far IR or UV, such as the gauge boson scattering amplitudes discussed above.

\subsection{Comments on real world QCD}

Much of the original work on hidden local symmetries was motivated by trying to understand the chiral Lagrangian of low energy QCD\@.  Indeed, this was also the primary phenomenological focus of refs.~\cite{Komargodski:2010mc,Kitano:2011zk}.  Consequently, our discussion would not really be complete without some brief comments on lessons that might be learned in this area.

Real world QCD is not supersymmetric so there is a limit to how trustworthy any insights derived from Seiberg duality can be.  In particular, many of our conclusions rely on the existence of quasi-NGBs that simply do not exist in the absence of SUSY\@.  One could, for example, attempt to construct a chiral Lagrangian accounting for heavy quarks by writing down an electric theory with $N=n=3$.  Three flavours (top, bottom and charm) have masses way above the electric (i.e.\ QCD) scale so are integrated out.  The electric theory subsequently confines at a scale $\Lambda_{\rm QCD}^6=m_tm_bm_c\Lambda_{\rm el}^3$.

Since $N=n$ for this theory the magnetic description has the same gauge group.  In addition, section \ref{sec:mu} suggests $\mu=\Lambda_{\rm el}=\Lambda_{\rm mg}$ for such theories.  The gauge group is higgsed to $SU(2)$ at a scale $\sqrt{\mu m_t}$, which is further higgsed to nothing at a scale $\sqrt{\mu m_b}$.  We therefore have magnetic gauge fields with masses
\begin{align}
m_3 & =g_{\rm mg}\pfrac{m_t^{1/3}\Lambda_{\rm QCD}}{(m_bm_c)^{1/6}}\frac{m_t^{1/3}\Lambda_{\rm QCD}}{(m_bm_c)^{1/6}} &
m_2 & =g_{\rm mg}\pfrac{m_b^{1/3}\Lambda_{\rm QCD}}{(m_tm_c)^{1/6}}\frac{m_b^{1/3}\Lambda_{\rm QCD}}{(m_tm_c)^{1/6}} 
\end{align}
for running gauge coupling $g_{\rm mg}$, evaluated at each of the two higgsing scales.  A QCD scale of $\Lambda_{\rm QCD}=400$ MeV would give $m_3\approx4$ GeV and $m_2\approx0.7$ GeV: too small to correspond to any of the vector mesons in the heavy quark sector of QCD\@.  However, this is to be expected.  The magnetic gauge fields arise from quasi-NGBs in SQCD, which have no right to remain massless without SUSY\@.  It is therefore unsurprising that they end up with much larger masses in QCD, where SUSY is taken away.

An aspect of the above discussion {\em not} strongly affected by SUSY is the argument used to fix the value of $a$ in section \ref{sec:rho}.  The general strategy was to derive two separate expressions for the magnetic gauge fields: one from the equations of motion turned out by the HLS formalism, and one from the Noether currents for the unbroken flavour symmetry.  By comparing them we were able to fix the normalisation of the magnetic quarks and, consequently, the value of $a$.

The same idea ought to work for real world QCD\@.  In SQCD we knew exactly what the electric and magnetic quark expectations were, and could easily write down explicit expressions for the NGBs\@.  In QCD the moduli space is less well understood, with fermion, rather than scalar, expectations providing the order parameters for the symmetry breaking.  Hence a direct translation of our results is not possible.

Nonetheless, the HLS formalism has been applied to massless, two flavour QCD \cite{Bando:1984ej,Bando:1987br} where the quark condensate is assumed to break the chiral $SU(2)_L\times SU(2)_R$ flavour symmetry to its diagonal subgroup.  The formalism results in HLS gauge field equations of motion that set
\be
V^\alpha_\mu=-\epsilon^{\alpha\beta\gamma}\pi^\beta\partial_\mu\pi^\gamma
\ee
for pions $\pi^a$.  One can also write down the current for the unbroken flavour symmetry in the HLS description.  In the unitary gauge it is \cite{Komargodski:2010mc}\footnote{In deriving this expression, the $SU(2)$ generators are chosen to satisfy $\Tr{S^\alpha S^\beta}=2\delta^{\alpha\beta}$.}
\be\label{eq:APJ}
J^\alpha_\mu=2av^2V^\alpha_\mu+2v^2(a-2)\epsilon^{\alpha\beta\gamma}\pi^\beta\partial_\mu\pi^\gamma+\mbox{3 particles.}
\ee

If we then consider a pure gauge field state the current is given by $J^\alpha_\mu=2av^2V^\alpha_\mu+\mbox{3 particles}$.  Substituting in the gauge field equations of motion simplifies the expression to $J^\alpha_\mu=-2av^2\epsilon^{\alpha\beta\gamma}\pi^\beta\partial_\mu\pi^\gamma+\mbox{3 particles}$.  On the other hand, one can calculate the current directly from the sigma model description to find $J^\alpha_\mu=-4\epsilon^{\alpha\beta\gamma}\pi^\beta\partial_\mu\pi^\gamma$.  Consistency thus requires $a=2$.

Of course, there remains the question of how a pure gauge field state can exist when the equations of motion fix $V^\alpha_\mu=-\epsilon^{\alpha\beta\gamma}\pi^\beta\partial_\mu\pi^\gamma$.  In a sense it cannot.  But as far as the physical currents are concerned it can, provided $a=2$ such that the pion contribution disappears.  So $a=2$ can be considered a direct consequence of allowing `pure' gauge field states (i.e.\ $\rho$-mesons) to exist in the HLS description of QCD\@.

\subsection{On non-supersymmetric duality}

This discussion brings us finally to non-supersymmetric dualities which, of course, it would be very interesting to establish.   In the past there have been various attempts in this direction.  None of them are quite as compelling as Seiberg duality itself, mainly because non-supersymmetric theories do not usually have interesting moduli spaces.  Other matching tests such as 't~Hooft anomaly matching are significantly weaker, often not uniquely pinning down the dual description.  Moreover the most constraining anomalies involve the $R$-symmetry, which is not available in non-supersymmetric theories. 

We have seen that the notion of an HLS provides a somewhat more mechanical route to dualities.  To establish a pair of Seiberg duals one begins with the electric theory of interest. Its higgsing leads to a nonlinear sigma model that can in turn be expressed as a linearised HLS theory. The existence of $R$-symmetry then guarantees a modulus along which the gauge symmetries in both descriptions are smoothly restored.  This procedure does not rely on the usual tests of moduli space or anomaly matching. 

In principle at least, it is clear what would be required in order to establish such a duality without SUSY\@.  First one requires an electric theory that can be higgsed or confined to give a nonlinear sigma model.  This is linearised in the HLS formalism, and the identification of magnetic degrees of freedom made in the same manner as for the SUSY theories.  In order to be able to take a symmetry restoring limit one would then need a scaling symmetry to be present. Thus the theories of interest would most likely be of the kind discussed in \cite{Shaposhnikov:2008xi}, in which scale invariance is spontaneously broken by a ``dilaton'' playing much the same role as the conformal compensator (equivalently $\kappa_R$) in the SUSY case.

Having established a candidate duality in this way one could, of course, still apply all the usual tests.  Any duality constructed via the HLS formalism that passed them all would surely be on firm footing indeed.

\subsubsection*{Acknowledgements} We would like to thank Matt Buican, Tony Gherghetta and Matthew McCullough for discussion and comments.  JB is supported by the Australian Research Council.

\providecommand{\href}[2]{#2}\begingroup\raggedright\endgroup

\end{document}